\documentclass[useAMS,usenatbib]{mn2e}
\usepackage{graphicx}
\usepackage{amssymb}

\def\bq{\mbox{\boldmath $q$}}
\def\bp{\mbox{\boldmath $p$}}

\def\bQ{\mbox{\boldmath $Q$}}
\def\bP{\mbox{\boldmath $P$}}
\def\mymathcalH{}
\def\HKep{\mymathcalH{H}_{\mathrm{Kep}}}
\def\Hint{\mymathcalH{H}_{\mathrm{int}}}
\def\HKepi{\mymathcalH{H}_{\mathrm{Kep}, i}}
\def\Hinti{\mymathcalH{H}_{\mathrm{int}, i}}
\def\HSun{\mymathcalH{H}_{\mathrm{Sun}}}

\def\Hij{\mymathcalH{H}_{i, j}}
\def\dd{{\partial}}
\def\aj{{\it AJ}}
\def\aap{{\it Ast.~\&Astrophys.}}
\def\apj{{\it Ap.~J.}}
\def\apjl{{\it Ap.~J.~Lett.}}
\def\mnras{{\it Mon}. {\it Not}. {\it R}. {\it Astr}. {\it Soc}.,~}
\def\Earth{\earth}
\def\ME{$\mathrm{M_{\Earth}}\,$}

\title[Methods for large dynamical range problems]{New methods for large
dynamical range problems in planetary formation} \author[D.~S.~McNeil
and R.~P.~Nelson]{D.~S.~McNeil$^{1}$\thanks{E-mail:
d.mcneil@qmul.ac.uk} and R.~P.~Nelson$^{1}$\\
$^{1}$Astronomy Unit, School of Mathematical Sciences, Queen Mary,
University of London, Mile End Road, London, UK E1 4NS}
\begin{document}

\date{Accepted 2008 October 19.  Received 2008 April 1.}

\pagerange{\pageref{firstpage}--\pageref{lastpage}} \pubyear{2007}

\maketitle

\label{firstpage}

\begin{abstract}

Modern N-body techniques for planetary dynamics are generally based on
symplectic algorithms specially adapted to the Kepler problem.  These
methods have proven very useful in studying planet formation, but
typically require the timestep for all objects to be set to a small
fraction of the orbital period of the innermost body.  This
computational expense can be prohibitive for even moderate particle
number for many physically interesting scenarios, such as recent
models of the formation of hot exoplanets, in which the semimajor axis
of possible progenitors can vary by orders of magnitude.  We present
new methods which retain most of the benefits of the standard
symplectic integrators but allow for radial zones with distinct
timesteps.  These approaches should make simulations of planetary
accretion with large dynamical range tractable.  As proof-of-concept we
present preliminary science results from an implementation of the
algorithm as applied to an oligarchic migration scenario for forming
hot Neptunes.

\end{abstract}


\section{Introduction}
\label{section:intro}

Dynamical timescales in the solar system vary widely.  For example,
Mercury's orbital period is 0.24 yr, Pluto's is 250 yr, and comets in
the Oort cloud can have periods of $\sim\!10$ Myr, for a difference of
over seven orders of magnitude.  The disparity in timescale increases
if we also consider not merely orbital period but period at periapse,
such as in Sun-grazing comets which can come within a few solar radii,
or the orbits of moons and satellites.  The age of our solar system is
$\sim\!4.6$ Gyr, and relevant formation timescales are believed to be
on the order of a few million years (for the accretion of giant planet
cores) to tens of millions of years (for the formation of the Earth)
to hundreds of millions of years (for possible late-stage
rearrangement of the outer solar system, e.g.~\citealt*{nice1}).

The enormous number of orbits required presents a considerable
challenge for numerical studies of planet formation.  In N-body
studies of galaxy formation, by contrast, the number of dynamical
times is low, and therefore the emphasis has been on increasingly
complex modelling of the physics and ever-larger numbers of particles,
both of which are amenable to parallelization.  Although
multiprocessor codes can offer major benefits even for planetary
simulations, the large number of timesteps limits the particle number
to a regime where latency issues loom large.  The planetary problem is
simply, and inescapably, hard.

The difficulties are yet greater for studies of the formation of hot
Neptunes, giant planets orbiting very close to the parent stars.
These planets are unlikely to have formed in situ, suggesting that gas
disc induced migration will play an important role.  For example,
GJ436b is a hot transiting Neptune \citep{butler2004} whose density is
$\sim\!1.69\,g/cm^3$ \citep{torres07}.  This suggests it is a true
Neptune analogue (i.e.~an ice giant, not a small gas giant or a large,
rocky super-Earth), which raises the question of where its ice
originated.  In the standard models of the minimum-mass solar nebula,
the snow line beyond which ices can condense is $\sim\!2.7$ AU.
Therefore one obvious toy scenario for the history of GJ436b is that
the planet started its life in the outer regions of the disc, and then
migrated in due to interactions with the gas, possibly growing en
route.

At present, the most elegantly constructed general integrator for
doing late-stage N-body studies of planet formation is SyMBA
\citep{dll98}, a Kepler-adapted symplectic integrator capable of
resolving close encounters, which is descended from the original
methods of \cite{wishol} and \cite{kyn}.  There are several
implementations of the algorithm and its variants available, including
a parallel version which has been useful in studying terrestrial
accretion \citep{mcn05}, and so it would be natural to apply these
codes immediately.  Unfortunately, a direct treatment of the hot
Neptune problem is completely beyond the reach of standard methods, at
least at the usual resolution.  The orbital period at 0.05 AU is
$\sim\!0.01$ yr, 30 times smaller than at 0.5 AU (a respectable inner
boundary for studies of the formation of the Earth).  Since migration
over several AU plays a role, we cannot concentrate on a narrow region
(e.g.~\citealt*{ko98}) but must build a more global model, requiring
large numbers of particles.  Furthermore, since the formation and
migration timescales will be important and often comparable, some
common tricks for speeding up simulations (such as increasing the
collisional radius) may be dangerous.

These formation scenarios are of considerable interest, and are not
easily studied using current numerical and computational technology;
they are messy systems, with many non-Hamiltonian forces and events,
and not clean celestial mechanics problems; and there seems to be a
promising direction for improvement.  We therefore seek to develop new
integrators which will allow us to address these problems.  Given the
difficulties, we are willing to consider approximations which we would
hesitate to use in other situations, such as a detailed study of
long-term chaos in the outer solar system.  We will instead sacrifice
some precision while preserving reliability in the hopes of exploring
otherwise inaccessible science: in this situation practicality beats
purity.

Fortunately, the very feature which makes the problem so challenging
-- the wide dynamical range -- opens up possibilities for new methods.
SyMBA and its cousins use a common timestep for all (non-encountering)
objects which is set by the minimum pericentric distance.  This
limitation can be worked around when objects only occasionally enter
the innermost regions \citep{ld00} but it reduces to using
unacceptably slow Bulirsch-Stoer integration when there is always an
object in the innermost regions, as we expect in our models.  However,
we note that for an inner edge of 0.05 AU, a timestep of $\sim\!0.0005$
yr would be necessary, but if the inner edge of the problem were 1 AU,
a timestep of 0.05 yr would suffice.  This suggests that using the
common small timestep results in objects beyond 1 AU being
`over-integrated' by roughly a factor of 100.  If we could somehow use
the larger timestep for the outer objects, then since most particles
in simulations of oligarchy tend to be in the outer regions (where
formation times are longer), we might be able to recover something
like the standard run times.  Indeed, a decade and a half ago,
\cite{saha94} were already building mixed-variable integrators with
different timesteps associated with each planet, so there is
precedent.

Therefore we set out to construct a new multizone method using recent
numerical technology which allowed for small timesteps in the inner
regions and large timesteps in the outer regions.  We had several
desired properties:

(1) Symplecticity, or at least near-symplecticity, is highly
    desirable.  By contrast, time-reversibility (in the absence of
    mergers and dissipational forces) is necessary, as many of the
    good conservation properties of symplectic integrators are
    inherited from their reversibility.

(2) SyMBA's underlying integrator step is very robust, especially at
    low eccentricities, and has extensive field-testing (both in
    Duncan and Levison's SWIFT and in John Chambers's MERCURY). An
    integrator which reduces to this proven approach for objects which
    are in the same zone is preferable.

(3) Correct close encounter handling is vital.  Although we may be
    willing to accept a decrease in encounter accuracy in a few
    locations (such as at the zone boundaries), the vast majority of
    encounters must be treated using a method known to be reliable.
    More generally, force inaccuracies should be limited to distant
    interactions, as in fast force techniques such as treecodes.

(4) Any discontinuities caused by the existence of distinct timestep
    zones should be kept to a minimum, and at or below second order if
    possible.

By using existing techniques in the literature, we develop integrators
which meet the above criteria, choosing at every branch point the
simplest scheme which seems likely to work.  We combine the
Hamiltonian splitting of \cite{dll98} with a multistep approach
inspired by \cite{saha94}, use the transitioning approach of
\cite{cham99} to preserve symplectic behaviour, and derive new
transition functions to make the scheme sufficiently smooth.

In \S\ref{section:sym} we briefly review the use of symplectic methods
in planetary dynamics.  In \S\ref{section:mvs} we introduce the
Kepler-adapted mixed-variable symplectic integrators; in
\S\ref{section:close_enc} and \S\ref{section:close_enc_with_Sun} we
explain methods for treating close encounters between planets and
between a planet and the Sun, respectively; and in
\S\ref{section:ind_timestep} we discuss integrators which allow
individual timesteps.  In \S\ref{section:naoko1} we construct the new
integrators, and in \S\ref{section:trans} we develop appropriate transition
functions.  We presents tests of the method in \S\ref{section:tests}
and a discussion in \S\ref{section:disc}.  We conclude in
\S\ref{section:conc}.

\section[]{Symplectic methods for Keplerian potentials}
\label{section:sym}

Geometric integrators are numerical integration methods which attempt
to build the properties of the equations and their solutions into the
integrators themselves, properties such as symmetries and their
corresponding conserved quantities (\citealt*{yosh93}).  This matching
of geometry between the problem and the solver can lead to
improvements in accuracy and robustness, as well as dramatic increases
in speed.  Symplectic integrators are geometric integrators where the
geometry of interest is Hamiltonian, and the conserved quantity is the
natural area element, the Poincar\'{e} 2-form $d \bp_i \wedge d
\bq^i$, where $\bq$ and $\bp$ are the usual generalized positions and
momenta, respectively, with summation over particles $i$ with mass
$m_i$.)  A symplectic integrator can be constructed via operator
methods or frequency maps; we restrict ourselves to operators.

For some Hamiltonian $\mymathcalH{H}$, if we divide it (arbitrarily but
conveniently) into a kinetic term $\mymathcalH{H}_{\mathrm{T}}$ and a
potential term $\mymathcalH{H}_{\mathrm{V}}$,
\begin{equation}
\mymathcalH{H} = \mymathcalH{H}_{\mathrm{T}} + \mymathcalH{H}_{\mathrm{V}}
\end{equation}
we can approximate the evolution under $\mymathcalH{H}$ by a second-order
time-reversible drift-kick-drift `leapfrog' scheme
\begin{equation}
\mymathcalH{H}^{\,\tau} \approx \mymathcalH{H}_{\mathrm{T}}^{\,\tau/2} \,\,
\mymathcalH{H}_{\mathrm{V}}^{\,\tau} \,\,
\mymathcalH{H}_{\mathrm{T}}^{\,\tau/2}
\end{equation}
where $H_X^{\, \tau}$ is the operator generated by the Hamiltonian
term $\mymathcalH{H}_X$, and $\tau$ is the timestep.  This integrator
solves a nearby `surrogate' Hamiltonian $\widetilde{\mymathcalH{H}} =
\mymathcalH{H} + \mymathcalH{H}_{\mathrm{err}}$ where
$O(\mymathcalH{H}_{\mathrm{err}}) = \tau^2$, using `solves' in the sense of
physics and not mathematics.  ($\mymathcalH{H}_{\mathrm{err}}$ is a purely
formal series which need not converge in general, and certainly need
not converge at the large timesteps used in practice, but is
well-approximated by its first few terms.)

\subsection{Wisdom-Holman method}
\label{section:mvs}

The work of \cite{wishol} and \cite{kyn} sparked a dramatic revolution
in planetary integrations by specializing the general techniques of
symplectic integration to the unique properties of Keplerian dynamics
in the solar system.  It is well known that the orbits of a system of
small bodies around a large central mass are very nearly conic
sections, and this Keplerian motion can be advanced (almost)
analytically.  This suggests that instead of dividing the Hamiltonian
into kinetic and potential terms as in the standard leapfrog, we
should divide it into a Keplerian term and a term corresponding to the
perturbations between the planets:
\begin{equation}
\mymathcalH{H} = \mymathcalH{H}_{\mathrm{Kep}} + \mymathcalH{H}_{\mathrm{int}}
\end{equation}
The difficulty arises in finding a canonical set of coordinates in
which the N-body Hamiltonian takes this shape.  \cite{wishol}
discovered that using Jacobi coordinates succeeds.  In this coordinate
system, the position and momentum of the $j$th object (of $N$ planets,
where the masses are given by the $m_j$, and the Sun is $j=0$) are
given relative to the centre of mass of the inner bodies, i.e.~those
with index $k < j$.  They write
\begin{equation}
\label{eq:Jacobi_HKep}
\HKep = \mathop{\sum}_{j=1}^{N} \left( \frac{|{\bp'_{j}} \cdot {\bp'_{j}}|}{2 m'_j} - 
\frac{G \, m_j \, m_0}{|\bq'_j|} \right )
\end{equation}
and
\begin{equation}
\label{eq:Jacobi_Hint}
\Hint =\mathop{\sum}_{j=1}^{N} \left ( 
  \frac{G \,\! m_j \,\! m_0}{|\bq'_j|} -
  \frac{G \,\! m_j \,\! m_0}{\bq_{j0}} \right ) - 
  \mathop{\sum}_{j=1}^{N-1}\mathop{\sum}_{k=j+1}^{N} 
  \frac{G \,\! m_j \,\! m_k}{\bq_{jk}}
\end{equation}
where the primed quantities are Jacobi coordinates and $\bq_{jk} =
\arrowvert \bq_{j} - \bq_{k} \arrowvert$.  In the absence of close
encounters, $\Hint \ll \HKep$ and the non-Keplerian perturbations on
the orbits due to mutual interactions are small.  We can construct a
second-order integrator as before, where the new $\HKep$ drifts are
`rolls' along the Keplerian conic section, and the $\Hint$ kicks are
the Cartesian perturbations between the planets; hence the name
`mixed-variable symplectic' (MVS; \citealt*{saha92}) has been used for
integrators of this type, as the integrations are in effect carried
through in both Cartesian and Keplerian variables.  This approximation
to the orbit is vastly superior to the linear-path kinetic/potential
decomposition, and the symplecticity provides robustness and
stability.  As a result, one can use much larger timesteps than would
be otherwise permissible, making planetary simulations running for the
age of the solar system feasible (e.g.~\citealt*{budd95}).

For a brief review of the history of various mapping methods in solar
 system dynamics, see ch.~9 of \cite{murray}.  \cite{wis06} also provides
 a useful review.

\subsection{Close encounters between planets}
\label{section:close_enc}

Despite its many benefits for studies of well-separated planets for
long timescales, the use of Jacobi coordinates in the MVS approach
limits its applicability.  Studies of planetesimal accretion, a highly
chaotic and stochastic process, require the ability to handle radial
reordering of objects and to resolve close encounters.  In the
Wisdom-Holman mapping, the coordinate frame depends upon a fixed
ordering of the objects, and cannot be updated as the system changes
without breaking symplecticity.  There is also no natural way to treat
encounters: first, by varying the timestep you lose the symplecticity
of the integration as the composition of two symplectic steps need not
be symplectic; and second, even if one could vary the timestep, the
coupling between objects in the Jacobi scheme means that all external
objects are also implicitly involved, so an encounter cannot be
treated independently.  We would prefer an integrator in which each
planet was treated equivalently and which could adapt the timestep
when planets strongly interact.

After some early experiments (e.g.~\citealt{ld94}), a superior
solution to the problem was presented in \cite{dll98} (hereafter
DLL98) which used an ingenious choice of coordinates to avoid the
coupling between planets, a particular choice of Hamiltonian splitting
to ensure that all planets were treated equally, and a clever
decomposition of the potential into shells to allow effective changes
in the timestep.  The authors developed a `democratic heliocentric'
(DH) method , where `democratic' means `symmetric with respect to the
labelling of the planets', and which avoids the particle entanglement
of Jacobi coordinates.  In these coordinates, one uses heliocentric
positions but barycentric momenta.  (Note that the same system is
called `mixed-centre' by \citealt*{cham99} and `canonical
heliocentric' by \citealt*{wis06}.)\footnote{The temptation to suggest
yet another name is resisted only with difficulty.}  To be explicit,
from $q$ and $p$ they compute new conjugate coordinates $\bQ$ and
$\bP$, such that

\begin{eqnarray}
\bQ_{i} = \left\{ \begin{array}{ll}
\frac{1}{m_{\mathrm{tot}}} \, \sum_{j=0}^{N} m_j {\bq}_j  & i = 0\\
\bq_i - \bq_0 & i \neq 0\\
\end{array}
\right.
\end{eqnarray}

and

\begin{eqnarray}
\bP_i = \left\{
\begin{array}{ll}
       \sum_{j=0}^{N} \bp_j  & i = 0\\
        \bp_i - \frac{m_i}{m_{\mathrm{tot}}} \, \sum_{j=0}^{N} \bp_j  & i \neq 0\\
\end{array}
\right.
\end{eqnarray}

This results in a new form for the Hamiltonian,
\begin{equation}
\label{dh_hamil}
\mymathcalH{H}(\bQ_i, \bP_i) = \HSun + \HKep + \Hint
\end{equation}
where
\begin{equation}
\HSun = \frac{1}{2m_0} \left\vert \sum_{i=1}^n \bP_i \right\vert^2,
\label{eq32b}
\end{equation}
\begin{equation}
\HKep = \sum_{i=1}^{n} \left( \frac{\vert\bP_i\vert^2}{2m_i} -
\frac {Gm_im_0} {\vert\bQ_i\vert} \right),
\label{eq32a}
\end{equation}

\begin{equation}
\Hint = -\sum_{i=1}^{n-1} \sum_{j=i+1}^{n} \frac{Gm_im_j}{
\vert\bQ_i-\bQ_j\vert}.
\end{equation}

$\HSun$ generates a linear drift of particles' positions (taking its
name from the fact it is determined by the barycentric momentum of the
Sun), $\HKep$ corresponds to a pure Kepler orbit, and $\Hint$, as
before, is the term due to the interaction of the particles.  Note
that -- unlike with the Wisdom-Holman mapping
(eqs. \ref{eq:Jacobi_HKep} and \ref{eq:Jacobi_Hint}) -- an encounter
between two particles is separable, in that the terms involving the
positions and momenta of the two objects can be pulled out of $\HKep$
and $\Hint$.  (In contrast, $\HSun$ is not separable in this way; we
return to this subject.)

As \cite{wis06} notes, the same canonical heliocentric coordinate
system was used in previous work \citep{toumawis93, toumawis94}, with
different Hamiltonian splittings.  For example, \cite{toumawis93} used
the alternate splitting
\begin{equation}
\HSun' = \frac{1}{m_0}  \sum_{1 \leq i < j}^n \bP_i \bP_j
\end{equation}
\begin{equation}
\HKep' = \sum_{i=1}^{n} \left( \frac{\vert\bP_i\vert^2}{2\mu_i} -
\frac {Gm_im_0} {\vert\bQ_i\vert} \right),
\end{equation}
where $\mu$ is the reduced mass.  This splitting has the advantage
that it preserves Kepler's semimajor axis-period relationship.
However, the use of the reduced mass means that two objects with the
same position and velocity but different masses will experience
different Kepler drifts and different linear drifts, which can present
difficulties for treating close encounters (DLL98, \S4).  In this
sense the DLL98 splitting is more `democratic', at the cost of being
slightly less accurate for a given orbit.  Accordingly, we will use
`canonical heliocentric' for the coordinate system itself, and reserve
`democratic heliocentric' for the specific three-term Hamiltonian
splitting used in DLL98 and described by eq.~\ref{dh_hamil}.

From this DH splitting, they construct a second-order integrator which
will be the basic DH step.  Following the conventional notation, let
$L$ be the operator generated by $\HSun$ (L for `linear drift'), let
$K$ correspond to $\Hint$ (K for `kick'), and let $D$ correspond to
$\HKep$ (D for `drift').  Using this form, the system is advanced one
timestep $\tau$ by applying
\begin{equation}
\label{eq:dh_basic_5_op}
H^\tau \approx L^{\tau/2} \, K^{\tau/2} \, D^{\tau} \, K^{\tau/2} \, L^{\tau/2}
\end{equation}

As naive adaptive timestepping changes the surrogate Hamiltonian and
thus breaks symplecticity, DLL98 develop a technique involving
recursive subdivision of the timesteps along with decomposing the
force into a series of shells around the particles, and associating
each force shell with a different timestep.  Similar approaches had
been attempted previously in molecular dynamics \citep{sk94} with
limited success, but DLL98 realized that there were relevant
smoothness constraints on the transition from shell to shell.

Starting from the basic DH step, we relabel $K,D$ as $K_0, D_0$.  We
replace the $D^{\tau}$ by a sequence of $M_i$ smaller substeps, $[
K_i^{\tau/2 M_i} \, D_i^{\tau/M_i} \, K_i^{\tau/2 M_i} ]^{M_i}$ where
$K_i$ is an interaction term to be defined, and then repeat the
process indefinitely at higher index:
\begin{eqnarray}
\begin{array}{ll}
\label{symba}
H &\approx L^{\tau/2} \, K_0^{\tau/2} \, D_0^{\tau} \, K_0^{\tau/2} \, L^{\tau/2}  \\
  &\approx L^{\tau/2} \, K_0^{\tau/2} [ K_1^{\tau/2M_1} \, D_1^{\tau/M_1}
\, K_1^{\tau/2M_1} ]^{M_1} \, K_0^{\tau/2} \, L^{\tau/2} \\
  &\approx L^{\tau/2} \, K_0^{\tau/2} [ K_1^{\tau/2M_1} \\
& \hspace{2.2cm} [ K_2^{\tau/2 M_1 M_2} \, D_2^{\tau/M_1 M_2} \, K_2^{\tau/2 M_1 M_2} ]^{M_2} \\
& \hspace{1.9cm} K_1^{\tau/2M_1} ]^{M_1} \, K_0^{\tau/2} \, L^{\tau/2} \\
&\approx \dots
\end{array}
\end{eqnarray}
Each of the above integrators, as well as the $i\to\infty$ limit, has
a fixed surrogate Hamiltonian; the timestep $\tau$ never changes.
Since the $D_i$ terms commute with themselves, then if succeeding
$K_i$ terms are merely the identity operators the neighbouring $D_i$
terms reduce to one large drift.  For example, if $K_i = 0$ for all $i
> 0$, then the integrator reduces to the basic DH step with timestep
$\tau$.  However, if $K_2$ were nonzero, then since $D_i$ and $K_i$ do
not commute, the reduction would not occur and the integrator would
act on the smaller timestep.  In this scheme, an infinite number of
terms are always active, but need not actually be considered unless
the intervening $K_i$ terms are nonzero.  The problem becomes finding
a decomposition of the potential $\Hint$ generating the $K$ terms
$K_0, K_1, K_2, \dots$ which will (1) recover the standard DH step
when no close encounters are occurring and the larger timestep $\tau$
suffices, i.e. $K_0=K$, and $K_i=0$ for $i > 0$; (2) always sum to the
correct amount of force, i.e. $\sum_{i} K_i = K$; and (3) do so in a
sufficiently smooth fashion.  To construct such a decomposition,
DLL98 imagine a decreasing sequence of shell radii $R_1 > R_2 >
\dots$ with a corresponding decomposition of the potential into $V_1,
V_2, \dots$ where the magnitude of the potential terms $V_k$ smoothly
decreases to zero outside the shell range.  By studying the error
Hamiltonian they find several necessary conditions on the
decomposition relating to the smoothness of the transition of the
potential from term to term.

The resulting algorithm SyMBA, combining the DH coordinate system with
the recursively subdivided smoothly-transitioning zone operators,
works remarkably well.  Nevertheless, it is challenging to implement
and difficult to test.  For example, a correct version of the
algorithm will often show worse energy and angular momentum
conservation on a given encounter than a subtly incorrect one,
requiring large test suites; and since the solution is very finely
tuned to the planetary problem, off-the-shelf routines are of little
use.

\cite{cham99} developed a much simpler though more expensive approach.
Of the three major advances in SyMBA -- DH coordinates, the smoothness
conditions on motion between operators, and the recursive subdivision
of the Hamiltonian -- only the first two are strictly necessary to the
resulting algorithm.  The major benefit of the recursion is that it
pushes the transitions entirely into the kick operators and preserves
the ability to apply the drift operator merely by solving Kepler's
equation, albeit at the cost of considerable complexity.  If we
surrender this requirement, however, then we can decompose the
Hamiltonian into three terms
\begin{equation}
\label{cham_eq}
\begin{array}{ll}
\mymathcalH{H} = \HSun + \left[ F(i,j) \, \Hij \right] + \left[ ( 1-F(i,j) ) \, \Hij + \HKep \right]
\end{array}
\end{equation}
where $\Hij$ is the component of $\Hint$ involving objects $i$ and
$j$, with implied summation over all $i,j$ with $i < j$. $F(i,j)$ is a
transition function which is 1 when the objects are distant and
approaches 0 when they are undergoing an encounter.  As in the basic
DH step (eq.~\ref{eq:dh_basic_5_op}), this decomposition gives rise to
a five-operator step, but these are the only five terms to consider,
unlike the much larger number of potentially active terms in
eq.~\ref{symba}.  It is true that whenever an encounter is occurring
and $F(i,j)$ could be nonzero then the interaction terms can no longer
be advanced analytically, but they can still be numerically integrated
to high precision using standard techniques such as Bulirsch-Stoer.
Moreover, and this advantage should not be underestimated, testing
that the above algorithm is implemented correctly is far more
straightforward than testing SyMBA.  Whenever the cost of the
numerical integration due to encounters is a small fraction of the
computation, the Chambers-style approach (although perhaps not as
elegant as SyMBA itself) is likely preferable on pragmatic grounds.
It should also be noted that despite the use of numerical integration
to handle the transition-weighted terms in the Hamiltonian, the
mapping itself remains symplectic; or as symplectic as any
floating-point implementation of sufficient precision can be.  (See
\citealt*{skeelapprox} for a general technique to recover
symplecticity when using otherwise non-symplectic approximations.)

\subsection{Close encounters with the Sun}
\label{section:close_enc_with_Sun}

In any case, both SyMBA and the Chambers variant have been
successfully applied to many studies of the later stages of planet
formation.  Their chief weakness is that neither can easily deal with
objects undergoing close encounters not with each other but with the
Sun, such as high-eccentricity Sun-grazing comets.  Unlike the case of
mutual encounters in which the important terms are $\Hint$ and
$\HKep$, during a close solar passage $\HSun$ must be evaluated more
frequently.  Therefore, to resolve such orbits correctly, one must
choose a timestep small enough to resolve the pericentre passage, and
that timestep must be fixed for the entire integration, even if such
encounters are very rare.  To overcome this limitation, \cite{ld00}
added a Chambers-style splitting on top of SyMBA (here we suppress the
planetary encounter terms, which are handled as described before) and
use
\begin{equation}
\begin{array}{ll}
\mymathcalH{H}^{\tau} = 
& (1-F) \, \HSun^{\tau/2} + \\ 
& \Hint^{\tau/2} + \\ 
& (F \, \HSun + \HKep )^{\tau} \, + \\ 
& \Hint^{\tau/2} + \\ 
& (1-F) \, \HSun^{\tau/2} \\ 
\end{array}
\end{equation}
\begin{equation}
\mymathcalH{H} = (1-F) \, \HSun  + 
 \Hint + 
 (F \, \HSun + \HKep ) 
\end{equation}
where $F$ is a transition function which is 0 when no object is near
the Sun and $1$ when {\em any} object is.  In this scheme, when any
object is undergoing a close solar approach, then the work of
performing an integration step is pushed into the new Kepler step
which is must be handled numerically.  The non-separability of $\HSun$
-- i.e.~the inability to isolate an individual object as in $\HKep$ --
becomes a serious inconvenience here, as it requires the numerical
integration of every object in the system even if only one object
enters the inner zone.  Nevertheless, as explained by \cite{ld00},
this approach provides the speed of SyMBA whenever the inner regions
are empty and yet can successfully survive occasional interlopers
(such as objects due to be ejected during violent periods in the
formation process).

This method cannot be directly applied as a solution for the numerical
challenge of hot exoplanet formation.  If we set the inner boundary at
a typical terrestrial-formation value like 0.5 AU, then there will
usually be many, and almost always be some, protoplanets and
planetesimals in the innermost region.  This means that every
particle's $\HKep$ will be numerically integrated on every step, and
the speed benefits are lost.  Moreover, if there are
mutually-gravitating objects inside the inner boundary, then the
situation is worse: the above integrator will evaluate the
interactions between bodies not undergoing close encounters on the
outer timestep $\tau$, which may bear little relation to the dynamical
times for the inner objects.  This particular problem could be
corrected by bringing $\Hint$ under the transition function $F$ as
well, but then we have merely recovered -- in an impressively
roundabout fashion -- a Bulirsch-Stoer integrator.

\subsection{Individual timesteps}
\label{section:ind_timestep}

Recognizing that in an MVS integration of the solar system, one is
taking hundreds of times more steps per Pluto orbit than would be
necessary if not for the presence of the interior planets,
\cite{saha94} construct a leapfrog integrator with individual
timesteps.  They split $\HKep$ and $\Hint$ into N terms each, such
that
\begin{equation}
\HKep = \sum_{i=1}^{N} {\HKepi} \,\, , \quad
\Hint = \sum_{i=1}^{N} {\Hinti} \,
\end{equation}
where $\HKepi$ is the Kepler term for planet $i$ (with objects
labelled in increasing semimajor axis), and $\Hinti$ is the
interaction term between planet $i$ and planets $i+1$ through $N$.
Translating into our notation, we use $D_i$ and $K_i$ to refer to the
operators as before, and assign a timestep $\tau_i$ to each planet,
where the largest timestep $\tau = \tau_N$ and $\tau_{i+1} / \tau_{i} =
M_i$ for $M_i$ an integer.  Starting with the Hamiltonian only
involving the outermost planet (note that $K_N$ is the identity
operator):
\begin{equation}
H^\tau \approx D_N^{\tau_N/2} K_N^{\tau_N} D_N^{\tau_N/2} \\
\end{equation}
and recursively applying
\begin{equation}
\label{eq:recur_substep}
K_i^\tau \rightarrow [D_{i-1}^{\tau_{i-1}/2} K_{i-1}^{\tau_{i-1}} K_i^{\tau}
D_{i-1}^{\tau_{i-1}/2}]^{M_{i-1}}
\end{equation}
for $i > 1$, one obtains an N-level integrator.  More concretely,
consider a two-planet case with timestep ratios of 1:3.  The resulting
integrator (after removing the $K_2$ term which does nothing)
\begin{equation}
H^\tau \approx D_2^{ \tau_2/2 }
[ D_1^{ \tau_1/2 }
K_1^{ \tau_1 }
D_1^{ \tau_1/2 } ]^{3}
D_2^{ \tau_2/2 }
\end{equation}
has a timestep of $\tau_1$ for the inner planet, and $\tau_2$ for the
outer planet, as desired.  Note that the integrator is time-reversible
even though the objects are not synchronized with respect to $D$ when
the interaction term $K_1$ is applied.  (Also note that in practice
one would combine neighbouring $D_1$ terms.)

It is important to recognize that the timesteps associated in this
method are individual but not adaptive; they must be set at the start
of the integration, and attach not to spatial zones but directly to
objects.  It is therefore unable to handle migrating objects, and as
an MVS method inherits the previously mentioned weaknesses of Jacobi
coordinates for our purposes.  However, it demonstrates that
asynchronous multi-stage integration in MVS-like contexts can be
constructed.

\subsection{Constructing the new integrator}
\label{section:naoko1}

We now have the necessary ingredients to construct a symplectic
integrator, Naoko (``New Adaptive Orthochronous Kepler Orbiter''),
which is Kepler-adapted, close-encountering, and yet allows for zones
with different timesteps.  Recall the basic DH step:
\begin{equation}
\label{dh_basic}
H^\tau \approx L^{\tau/2} \, K^{\tau/2} \, D^{\tau} \, K^{\tau/2} \, L^{\tau/2}
\end{equation}

We will seek a generalization of this step for the multiple-zone case.
We will define our timestep zones by dividing the system into radial
shells such that an object's instantaneous heliocentric radius
determines its zone assignment.  (This definition is mentioned here
for concreteness; other choices are possible.  We motivate this
particular choice in \S\ref{section:trans}.)  We label the zones using
integer indices, starting at 0 for the innermost zone, and an integer
subscript on an operator corresponds to the operator for that zone, in
a sense which will be made explicit later.  For example, $D_0^{\tau}$
advances all objects in zone 0 for a timestep $\tau$ under $\HKep$ but
does nothing to objects in other zones: the commutator bracket $[D_i,
D_j] = 0$.  Note that this differs from the \cite{saha94} usage of
subscripts to refer to planets.

To begin with, we defer consideration of interactions between the
planets.  Under this simplification, and starting with zone $1$,
eq.~\ref{dh_basic} becomes
\begin{equation}
H^\tau \approx L^{\tau/2} \, D_1^{\tau} \, L^{\tau/2} 
\end{equation}
The $L$ terms cannot be divided into zones as $\HSun$ is not
separable.  Nevertheless, we can incorporate more zones by applying
the individual-leapfrog approach of \cite{saha94} and recursively
subdividing the $L$ steps using this expression.  That is, we can
write
\begin{equation}
L^{\tau/2} \rightarrow  L^{\tau/4} \, D_0^{\tau/2} \, L^{\tau/4} 
\end{equation}
resulting in
\begin{equation}
\begin{array}{rc}
H^\tau \approx & \big[ L^{\tau/4 M_0} \, D_0^{\tau/2 M_0} \, L^{\tau/4 M_0} \big]^{M_0} \\ 
& D_1^{\tau} \\
& \big[ L^{\tau/4 M_0} \, D_0^{\tau/2 M_0} \, L^{\tau/4 M_0} \big]^{M_0}
\end{array}
\end{equation}
where $M_0$ is an integer setting the number of zone 0 steps per zone
1 step.  

This integrator looks promising.  In the absence of any objects in
zone 1, this is merely $2 M_0$ (kick-free) DH steps of size $\tau/2
M_0$ next to each other, and in the absence of any objects in zone 0,
the $D_0$ operators do nothing and the $L$ operators collapse,
reducing to a DH step of size $\tau$.  There are $2 M_0$ zone 0 drifts
per zone 1 drift, so if we assume each drift takes equal time then if
$N_1 > 2 M_0 N_0$, the zone 1 computation dominates.  As long as $N_1
\gg 2 M_0 N_0$, then handling objects in the innermost zone -- far
from requiring a major decrease in system timestep as in the standard
approach -- is effectively free; and, importantly, they can be handled
independently of the outer objects.

Despite appearances, the presence of the $L$ terms on the smallest
timescale does not remove this separability.  Although $L$ must be
formally {\em applied} with the innermost timestep, this does not mean
that $L$ must be {\em computed} at that frequency.  Since $D_0$ does
not affect objects in zone 1, and $L$ changes their positions but not
their velocities, we need only determine their contribution to $L$ at
the beginning of the zone 0 substep, and we can delay actually moving
them until the beginning of the zone 1 substep.  Implementing a
lazy-evaluation scheme is relatively simple.  (Since $L$ is so trivial
to evaluate and apply, it is seldom a bottleneck, at least in the
serial case.  For parallel implementations, we find that lazy
evaluation is vital, because otherwise the communication overhead
involved produces enormous scaling difficulties.  Potential
implementors should bear this warning in mind when designing data
structures.)

One can easily generalize to more zones.  The three-zone case is the
simplest integrator where the transition functions we will develop can
be carried through to a larger number of zones, and therefore we will
use it as our standard example.  In a minor abuse of notation, define
$S^\tau = L^{\tau/2} D^{\tau} L^{\tau/2}$ (note that unlike with $L$,
$D$, and $K$, $[S^\tau]^M \neq S^{M \tau}$.)

Then we have the two-zone integrator
\begin{equation}
\mymathcalH{H}^{\, \tau} = \big[S_0^{\tau/2 M_0}\big]^{M_0} 
\, D_1^{\tau} \,
\big[S_0^{\tau/2 M_0}\big]^{M_0} 
\end{equation}

and the three-zone version
\begin{eqnarray}
\begin{array}{rc}
\mymathcalH{H}^{\, \tau} = & \big[
\big[S_0^{\tau/4 M_0 M_1}\big]^{M_0} 
\, D_1^{\tau/2 M_1} \,
\big[S_0^{\tau/4 M_0 M_1}\big]^{M_0} 
\big]^{M_1} \\
& D_2^{\tau} \\
& \big[
\big[S_0^{\tau/4 M_0 M_1}\big]^{M_0} 
\, D_1^{\tau/2 M_1} \,
\big[S_0^{\tau/4 M_0 M_1}\big]^{M_0} 
\big]^{M_1} 
\end{array}
\end{eqnarray}
where $M_i$ sets the number of zone $i$ substeps per zone $i+1$ step.

After one step, objects in all zones have been advanced the correct
total time under each operator.  Moreover, neglecting the influence of
objects in other zones, each zone has experienced what is locally a
standard DH step: objects in zone 0 took a timestep of $\tau/4 M_0
M_1$; objects in zone 1 had a timestep of $\tau/2 M_1$; and objects in
zone 2 had a timestep of $\tau$.

We must now choose where to place the force operators.  To simplify
the discussion, we set $M_0 = M_1 = 1$, and assume that the zones are
separated such that $\tau$, $\tau/2$, and $\tau/4$ are appropriate DH
timesteps for all objects in the respective zones; recovering the
general case is straightforward.  The kick-free three-zone integrator
is given by

\begin{eqnarray}
\begin{array}{rc}
\mymathcalH{H}^{\, \tau} = & L^{\tau/8} 
D_0^{\tau/4}
L^{\tau/8} 
D_1^{\tau/2}
L^{\tau/8} 
D_0^{\tau/4}
L^{\tau/8} 
\, D_2^{\tau} \\
& L^{\tau/8} 
D_0^{\tau/4}
L^{\tau/8} 
D_1^{\tau/2}
L^{\tau/8} 
D_0^{\tau/4}
L^{\tau/8} \\
\end{array}
\end{eqnarray}

Let $K_{ij}$ be the kick operator between zones $i$ and $j$.  In order
to treat close encounters using the Chambers splitting, we must have
each $D_i$ surrounded by two $K_{ii}$.
\begin{equation}
\begin{array}{rc}
\mymathcalH{H}^{\, \tau} = & L^{\tau/8} 
K_{00}^{\tau/8}
D_0^{\tau/4}
K_{00}^{\tau/8}
L^{\tau/8} \\
& K_{11}^{\tau/4}
D_1^{\tau/2}
K_{11}^{\tau/4} \\
& L^{\tau/8} 
K_{00}^{\tau/8}
D_0^{\tau/4}
K_{00}^{\tau/8}
L^{\tau/8} \\
& K_{22}^{\tau/2}
D_2^{\tau}
K_{22}^{\tau/2} \\
& L^{\tau/8} 
K_{00}^{\tau/8}
D_0^{\tau/4}
K_{00}^{\tau/8}
L^{\tau/8}  \\
& K_{11}^{\tau/4}
D_1^{\tau/2}
K_{11}^{\tau/4} \\
& L^{\tau/8} 
K_{00}^{\tau/8}
D_0^{\tau/4}
K_{00}^{\tau/8}
L^{\tau/8} \\
\end{array}
\end{equation}
This integrator does not incorporate interzone forces, but intrazone
forces are evaluated on the appropriate timescale relative to the
drift timescale for objects in that zone.  One can therefore apply the
Chambers-style encounter handling between the sets of $K_{ii} D_i
K_{ii}$ as described in section \ref{section:close_enc} and we recover
the standard approach.  (In some situations we have found the $D K D$
operator splitting is superior, but for reasons involving
implementation details we will restrict ourselves to discussing the $K
D K$ version.)

There are many possible arrangements for force communication between
zones.  We optimistically choose the scheme requiring the fewest force
calculations, and update forces between zones $i$ and $j$ on the
outermost zone's timestep.  This choice results in
\begin{eqnarray}
\label{eq:alg3}
\begin{array}{rc}
\mymathcalH{H}^{\, \tau} = & L^{\tau/8} 
K_{00}^{\tau/8}
D_0^{\tau/4}
K_{00}^{\tau/8}
L^{\tau/8} \\
& K_{01}^{\tau/4}
K_{11}^{\tau/4}
D_1^{\tau/2}
K_{11}^{\tau/4} 
K_{01}^{\tau/4}\\
& L^{\tau/8} 
K_{00}^{\tau/8}
D_0^{\tau/4}
K_{00}^{\tau/8}
L^{\tau/8} \\
& K_{02}^{\tau/2}
K_{12}^{\tau/2}
K_{22}^{\tau/2}
D_2^{\tau}
K_{22}^{\tau/2}
K_{12}^{\tau/2}
K_{02}^{\tau/2}\\
& L^{\tau/8} 
K_{00}^{\tau/8}
D_0^{\tau/4}
K_{00}^{\tau/8}
L^{\tau/8}   \\
& K_{01}^{\tau/4}
K_{11}^{\tau/4}
D_1^{\tau/2}
K_{11}^{\tau/4} 
K_{01}^{\tau/4} \\
& L^{\tau/8} 
K_{00}^{\tau/8}
D_0^{\tau/4}
K_{00}^{\tau/8}
L^{\tau/8} 
\end{array}
\end{eqnarray}
This approach succeeds in separating the integration of zone $i$
objects from zone $j$ objects.  Objects in zone 2 need only have
drifts and kicks involving them evaluated on the timestep $\tau$;
objects in zone 1 on $\tau/2$; and zone $0$ on $\tau/4$; and as
already discussed $L$ is not a problem.  Note that the above
integrator was built to minimize the number of force operators, but
one could evaluate the cross-zone kicks ($K_{12}$, for example) more
frequently if desired.

The above integration technique obeys Newton's third law regarding
interparticle forces.  Although forces between objects in different
zones are computed less frequently than forces within a zone, whenever
gravitational accelerations are computed between two bodies the
accelerations are equal and opposite (even for close-encountering
objects in a transition zone, to be discussed later) and they are
immediately applied and turned into changes in velocity.  No force lag
or accumulation is involved.

\subsection{Transition functions}
\label{section:trans}
We recall that the key insight of DLL98 is that to preserve
symplectic behaviour while effectively changing the integration step
(and therefore the surrogate Hamiltonian) an object must experience a
smooth transition from one integration regime to another.  Roughly
speaking, if an object's transition is sufficiently smooth, then
instead of suddenly finding itself evolving under a different
surrogate Hamiltonian, it believes it is merely in a different regime
of the original Hamiltonian.  Accordingly, we now return to the
previously-deferred issue of choosing appropriate transition
functions, which turns out to be the most challenging part of the
problem.

\begin{figure}
\includegraphics[width=85mm,height=73.6667mm]{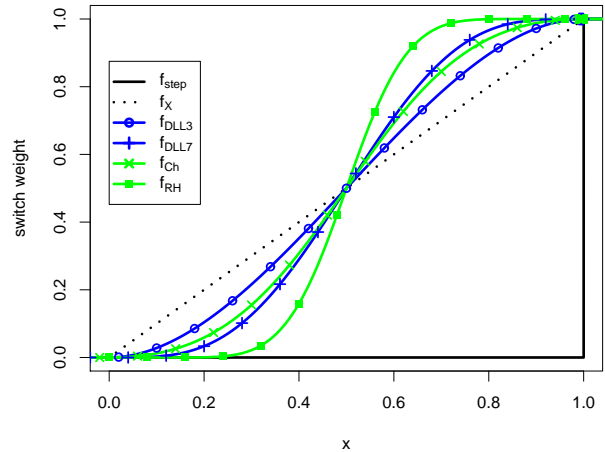}
\caption{Various candidate switch functions, discussed in \S\ref{section:trans}.}
\label{fig:trans}
\end{figure}

Let $f(x)$ be a switch function, a real-valued nondecreasing function
defined on $[0, 1]$ with $f(0)=0$ and $f(1)=1$.  We take the extension
outside this domain (equal to 0 below and 1 above) as given, under
which convention $f(x) = x$ is a switch.  The simplest switch is a
shifted step function:
\begin{equation}
f_{\mathrm{step}}(x) = \left\{\begin{array}{ll}
0 \quad & x < 1 \\
1 \quad & x \ge 1\\
\end{array}
\right.
\end{equation}
However, this leads to sudden movement of portions of the Hamiltonian
from one term to another, and symplecticity is lost.
DLL98 suggest 
\begin{equation}
f_{\mathrm{DLL3}}(x) = 3 x^2 - 2 x^3
\end{equation}
as one of their switches, which has $f'(x)=0$ at both endpoints, as
well as the higher-order
\begin{equation}
f_{\mathrm{DLL7}}(x) = x^4 (35 - 84 x + 70 x^2 - 20 x^3)
\end{equation}
\cite{cham99} suggests
\begin{equation}
f_{\mathrm{Ch}}(x) = x^2 / (2x^2 - 2x + 1)
\end{equation}
as a useful compromise between smoothness and
efficiency of evaluation.  \cite{rauch} prefer
\begin{equation}
f_{\mathrm{RH}}(x) = \frac{1}{2} \left( 1 + \tanh \left[ \frac{2x
-1}{x(1-x)}\right] \right)
\end{equation}
for which all derivatives vanish at the endpoints.  Figure
\ref{fig:trans} shows the various functions.  We will use the cubic
polynomial switch $f_{\mathrm{DLL3}}$, but the construction is
independent of this choice.  (This issue is discussed further in
\S\ref{section:vary_switch}.)

To reduce clutter we define a rescaling
function on the switch,
\begin{equation}
C(x, x_0, x_1) = f_{\mathrm{DLL}}\left(\frac{x-x_0}{x_1-x_0}\right)
\end{equation}
and build our transition functions from this base.

\begin{figure}
\includegraphics[width=85mm,height=73.6667mm]{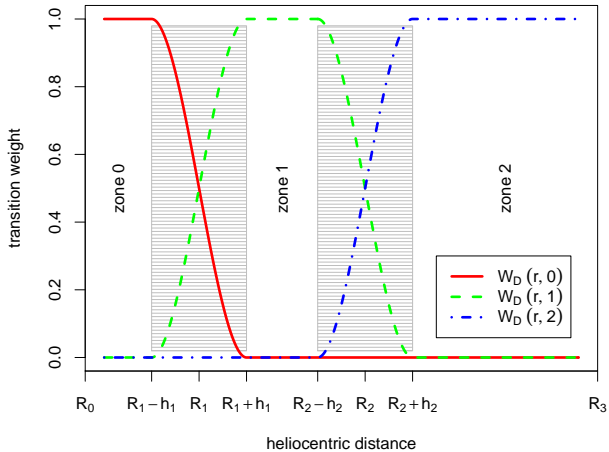}
\caption{A sketch of the radial zone scheme.
\label{schem1}}
\end{figure}

As described briefly in \S\ref{section:naoko1}, we divide the system
into zones based on instantaneous heliocentric radius $r$, and imagine
a set of spherical shells around the Sun.  Each zone has a transition
region near the interior and exterior edge of the shell in which
operators involving both adjacent zones will act on an object.  Let
$R_i$ be the locations of the $i$ zone boundaries and $2 h_i$ be the
widths of the transition regions, as illustrated in figure
\ref{schem1}.

An unfortunate consequence of this choice is that the resulting
evolution of transiting objects under $D$ is no longer analytically
integrable.  It is tempting to construct a transition function based
not on $r$ but on the osculating semimajor axis $a$, which is constant
during $D$ and can therefore be used (along with all other orbital
elements save those such as the mean anomaly describing the position
along the orbit) to build a weighted but integrable $D$.  Indeed,
experiments show that an $a$-based function can work for isolated
objects in the transition region.  However, doing so vastly
complicates the treatment of close encounters, as two encountering
objects must have similar instantaneous $r$ but can have very
different $a$: consider a high-$e$ zone 0 object at apocentre meeting
a high-$e$ zone 1 object at pericentre.  Under an $r$-based weighting
scheme, the difference in their effective drift timestep is bounded by
their physical radial separation which gets smaller as the encounter
gets deeper, whereas under an $a$-based scheme the drift timestep
experienced by the inner object could remain very different from that
of the outer object, which is not a recipe for numerical stability.  A
function which smoothly changes dependence from $r$ to $a$ with
changes in the encounter status may be possible, but in our view the
potential benefits are outweighed by the resulting complexity.

For the integrators presented here, such as that of eq.~\ref{eq:alg3},
there are three types of transitions we must consider: (1) those
involving drifts; (2) those involving distant kicks; and (3) those
involving close encounters.  Each has its own peculiarities, and
treating (2) and (3) simultaneously is rather awkward.

\subsubsection{Drift transitions}

The drift transitions are the easiest to handle.  We need a function
which will yield (for example) the full $D_1$ term for an object
completely within zone 1, and likewise with $D_2$ and an object in
zone 2, but which will return smoothly-varying intermediate values for
objects inside the transition zones $[R_1 - h_1, R_1 + h_1]$ and $[R_2
- h_2, R_2 + h_2]$, and generally $[R_i^{-}, R_i^{+}]$ where
$R_i^{\pm} \equiv R_i \pm h_i$.  (For simplicity we will restrict
ourselves to considering symmetric functions, although since objects
in inner zones will have higher velocities and different effective
timesteps, it is possible that an asymmetric function could yield
better results.)  Such a function is given by
\begin{equation}
\label{eq:W_D}
W_{D}(r, i)= \left\{
\begin{array}{ll} 
0 & r \le R_i^{-} \\
C(r, R_i^{-},R_i^{+}) & R_i^{-} \le r \le R_i^{+} \\
1 & R_i^{+} \le r \le R_{i+1}^{-} \\
1 - C(r, R_{i+1}^{-}, R_{i+1}^{+}) & R_{i+1}^{-} \le r \le R_{i+1}^{+} \\
0 & R_{i+1}^{+} \le r \\
\end{array}
\right.
\end{equation}
where $r$ is the instantaneous heliocentric radius and $i$ is the zone
index.  This is nothing more than a SyMBA-style transition applied to
$r$ instead of to the interplanet separation.  Although in general each
$R_i$ can have an associated transition zone, in practice the boundary
`transitions' are slightly degenerate.  Typically one would set the
inner transition region, $(R_0, h_0)$, and the outer transition region,
here $(R_3, h_3)$, well outside the radii of interest so that the sum
of weights over all zones is equal to 1 for all integrated objects
regardless of $r$.  In practice we make $R_0$ smaller than our inner
edge (set by the physics of the problem or the numerics of our
timestep), $R_3$ larger than the outer edge, and let $h_0$ and $h_3$
be some arbitrary small distance.  Figure \ref{schem1} sketches the
resulting scheme, with three timestep zones and effectively two (not
four) transition zones.

\subsubsection{Distant forces}

 Treating the forces is more difficult, as it involves not only the
interacting bodies' two distinct orbital radii but the separations
between them. We seek a function which will ensure the objects
experience sufficient continuity in their forces as they move from
zone to zone.  We will first consider only distant forces (i.e.~we
imagine all forces are `soft', requiring no special attention) and
then correct to handle close encounters.

Consider an integrator with three zones, labelled $i$, $j$, and $k$
from innermost to outermost, and two planets, with orbital radii $r_0$
and $r_1$.  Let the objects start in the zone $j$.  Initially, all the
weight should be in $K_{jj}$.  If we keep $r_0 = r_1$ and move the
pair together both inwards and outwards, we see that we need smooth
transitions between $K_{ii}$ and $K_{jj}$ as well as between $K_{jj}$
and $K_{kk}$.  If we instead keep $r_0$ constant while increasing
$r_1$, then as the second object crosses from $j$ to $k$, then the
$K_{jj}$ weight must decrease and $K_{jk}$ must increase until $r_1$
is fully within $k$ and the $K_{jk}$ weight is 1.  If we decrease
$r_0$ and increase $r_1$, then when both objects are in the centre of
the transition zones ($i$-$j$ and $j$-$k$), we will need to spread the
force out over four different terms: $K_{ij}$, $K_{jk}$, $K_{ik}$, and
$K_{jj}$.

This can be achieved by symmetrizing $W_D$:
\begin{equation}
W_{\mathrm{dist}}(r_0, r_1, i, j) = \left\{ 
\begin{array}{ll} 
(W_D(r_0, i) \, W_D(r_1, j) \, + & i = j \\ 
\quad\,W_D(r_1, i) \, W_D(r_0, j))/2 & \\
W_D(r_0, i) \, W_D(r_1, j) \, + & i \ne j \\ 
\quad\, W_D(r_1, i) \, W_D(r_0, j) & \\
\end{array}
\right. 
\end{equation}

\subsubsection{Close encounters}
Now we must ensure that close encounters are correctly shared.  This
requires a transition function $W_{\mathrm{cl}}$ which will move the force
between two objects from the $K_{ii}$ term to the $D_i$ term as in the
Chambers methods, resulting in a triplet of terms
\begin{equation}
\left[\frac{1}{2} W_{\mathrm{cl}} K_{ii} \right] + 
\left[(1 - W_{\mathrm{cl}}) K_{ii} + D_i \right] + 
\left[\frac{1}{2} W_{\mathrm{cl}} K_{ii} \right]
\end{equation}
eq.~\ref{cham_eq}.  We take
\begin{equation}
W_{\mathrm{cl}} = C(ds, ds_{\mathrm{crit}}/2, \, ds_{\mathrm{crit}})
\end{equation}
where $ds_{\mathrm{crit}}$, after DLL98, is several times the
sum of the Hill radii (with a possible additional dependence on
orbital velocity; as Chambers has noted, what matters is ensuring you
integrate through the transition.)  Note that unlike the transition
function recommended in \cite{cham99}, $W_{\mathrm{cl}}$ moves all of the
encounter into the numerically-integrated term $(1-W_{\mathrm{cl}}) K_{ii}
+ D_i$ at separation $ds = ds_{\mathrm{crit}}/2$, not at $ds = 0$.

\subsubsection{Combining the transitions}
Each of the three above transition functions -- $W_D$, which applies
to the D operators; $W_{\mathrm{dist}}$, which applies to all $K_{ij}$
operators; and $W_{\mathrm{cl}}$, which applies between $K_{ii}$ operators --
makes sense independently.  The natural way to combine them would give
(for example, suppressing indices on the transition functions):

\begin{equation}
\label{eq:default_close}
\begin{array}{l}
(W_{\mathrm{dist}} K_{23})^{\tau_3/2} \, \, (W_{\mathrm{dist}} W_{\mathrm{cl}} K_{33})^{\tau_3/2} \\
\left[ W_{\mathrm{dist}} (1 - W_{\mathrm{cl}}) K_{33} + W_D D_{33} \right] ^{\tau_3}  \\
(W_{\mathrm{dist}} W_{\mathrm{cl}} K_{33})^{\tau_3/2} \, \, (W_{\mathrm{dist}} K_{23})^{\tau_3/2} 
\end{array}
\end{equation}

The above approach has a minor problem, however.  Consider two objects
undergoing an encounter in a transition zone (say, the $i$-$j$
boundary.)  The above functions will attempt to share force across
three terms: $K_{ii}$, $K_{jj}$, and the cross-term $K_{ij}$.
However, the integrator is only built to treat the encounter using the
$K_{ii} D_i K_{ii}$ and $K_{jj} D_j K_{jj}$ substeps.  Any force that
the $K_{ij}$ operator is assigned is only sampled on the larger
timestep, and if the two objects are both in the middle of the
transition zone this can be as much as half the total force.  This
will result in a highly inaccurate integration.

How can this be repaired?  In the case of an encounter, $K_{ii}$ and
$K_{jj}$ should share all the force:

\begin{equation}
W_{\mathrm{near}}(r_0, r_1, i, j) = \left\{ 
\begin{array}{ll} 
(W_D(r_0, i) \, W_D(r_1, j) \, + & i = j \\ 
\quad\,W_D(r_1, i) \, W_D(r_0, j))/2 & \\
0  & i \ne j \\ 
\end{array}
\right. 
\end{equation}

In the absence of encounters, the above method
(eq.~\ref{eq:default_close}) should work.  Therefore we define yet
another transition function,
\begin{equation}
W_{\mathrm{shift}} = C(ds, ds_{\mathrm{crit}}, 2 \, ds_{\mathrm{crit}})
\end{equation}
and use it to smoothly interpolate between the no-encounter case when
all operators are involved and the encounter case when only the
intrazone operators share the force.  (Here we will require that
objects cannot undergo mutual encounters unless they are in
neighbouring zones.  Removing this limitation is possible but
unnecessary for our intended applications.)  Combining the above, we
write
\begin{equation}
\label{eq:kick_final}
W_K(i,j) = 
W_{\mathrm{shift}} W_{\mathrm{dist}} + (1 - W_{\mathrm{shift}}) W_{\mathrm{near}}
\end{equation}
and replace instances of $W_{\mathrm{dist}}$ with this encounter-corrected
expression.

\subsection{Assembling the integrator}

The final integrator is constructed by applying the drift transition
function (eq.~\ref{eq:W_D}) and the kick transition function
(eq.~\ref{eq:kick_final}) to the multilevel step of eq.~\ref{eq:alg3}.
The order of the substeps and the relationships between the $\tau_l$
are determined by the recursive subdivision of
eq.~\ref{eq:recur_substep} (as in eq.~\ref{eq:alg3}).  For each substep
of level $l$, the algorithm proceeds as follows:\\\\
(1) Do the linear drift: if $l=0$, apply $L_0$ for $\tau_0/2$.\\
(2) Apply the distant forces between $l$ and all interior levels:
apply $W_K(i,l) \, W_{\mathrm{cl}} \, K_{il}$ for all $i \le l$ for
$\tau_l/2$.\\
(3) Drift (including possible encounters): apply the
distant forces between $l$ and all interior levels: apply $W_K(l,l) \,
(1 - W_{\mathrm{cl}}) \, K_{ll} + D_l$ for $\tau_l$.  (Note that any
implementation would avoid doing this numerically when the kick term
was known to be zero because the objects are too well-separated.)\\
(4) Apply the distant forces between $l$ and all interior levels:
apply $W_K(i,l) \, W_{\mathrm{cl}} \, K_{il}$ for all $i \le l$ for $\tau_l/2$.\\
(5) Do the linear drift: if $l=0$, apply $L_0$ for $\tau_0/2$.\\

A moment's consideration confirms that as promised in
\S\ref{section:intro}, this is nothing more than the step subdivision of
\cite{saha94} applied to the Hamiltonian of DLL98 with the
close encounter handling of \cite{cham99}, with staggered force calculations,
and some new transition functions incorporated to ensure smoothness.  

In practice, we do not use the above scheme directly, but instead use
a `lower-level' scheme (motivated in part by \citealt*{rauch}) in
which we do not apply the transitions between terms in the Hamiltonian
but between the resulting $\dd q/\partial t$ and $\partial p/\partial
t$.  That is, given two pieces of the Hamiltonian $H_A$ and $H_B$ and
a transition function $f = f(q)$ between them, instead of starting
with the decomposition
\begin{equation}
\big[ f H_A \big] + \,\, \big[ (1-f) H_A + H_B \big]
\end{equation}
which produces
\begin{eqnarray}
\begin{array}{ll}
\frac{dq}{dt} = \frac{\dd}{\dd p} (f H_A) &, \, \frac{dp}{dt} = -\frac{\dd}{\dd q} (f H_A) \\
\frac{dq}{dt} = \frac{\dd}{\dd p} ( (1 - f) H_A + H_B) & ,
\frac{dp}{dt} = -\frac{\dd}{\dd q} ( (1 - f) H_A + H_B )
\end{array}
\end{eqnarray}
as the derivatives to be integrated, we write
\begin{eqnarray}
\begin{array}{ll}
\frac{dq}{dt} = f \, \frac{\dd H_A}{\dd p} \,\, &, \,\, 
\frac{dp}{dt} = -f \, \frac{\dd H_A}{\dd q} \\ 
\frac{dq}{dt} = (1-f) \, \frac{\dd H_A}{\dd p} + H_B \,\, &, \,\,
\frac{dp}{dt} = -(1-f) \, \frac{\dd H_A}{\dd q} + H_B
\end{array}
\end{eqnarray}
and apply the transition directly between the derivatives themselves.
This saves some computation (of the cross-terms $H \dd f/\dd q$, at
least; the $H \dd f/\dd p$ terms are all 0 because $f$ is not a
function of the momenta) and increases the effective smoothness of the
transition for a fixed $f$ because one does not lose a degree of
smoothness passing from $H$ to its derivatives.  It might be objected
that this means the resulting integration technique is no longer
strictly symplectic, but in fact one can construct a Hamiltonian
(admittedly somewhat artificial) for which an obviously symplectic
integrator generates exactly this algorithm.  However, even if this
were not the case and we were to view the integrator as merely
`near-symplectic', it remains time-reversible and works well in
practice, which agrees with the results of \cite{rauch} who applied
the transition functions in their SyMBA-style decomposition to the
forces and not the potential.  Of course, one may work with the
original Hamiltonian-level splitting if preferred.

One easily overlooked issue which is only apparent when considering a
global simulation is that in the SyMBA scheme the (effective) Hill
radii -- the Hill radii used in the close encounter criteria -- are
fixed at the start of the integration.  This is necessary to ensure
that the integration remains both symplectic and analytically soluble.
However, in cases where there is considerable inward migration, an
object will have an unnecessarily large close encounter criterion when
it arrives at the inner portion of the disc.  Similarly, an outward
migrating object could miss encounters.  The strict solution is to
bring the Hill radius dependence of the close encounter weight
function under the numerical integration; the lazy
not-quite-symplectic solution is to update the effective Hill radii on
some interval, at the risk of interfering with ongoing encounters.
All things being equal, if the system is messy and the number of
updates is low it is unlikely to make a statistically significant
difference.

\section{Tests of the method}
\label{section:tests}

The algorithm was implemented using the existing codebase of miranda
\citep{mcnphd} as a framework, and using a standard Bulirsch-Stoer
routine to handle the numerically integrated terms in the Hamiltonian.
Since the method reduces to the well-understood Chambers approach in
the one-zone limit, we will concentrate on testing the performance of
the multiple-zone aspects of the code.  We have confirmed by
comparison with miranda in both SyMBA and Bulirsch-Stoer modes that
the new code behaves as expected in the one-zone case.  Unless
otherwise specified, the numerical integration tolerances were set at
$10^{-16}$ (which in practice generates errors $\sim\!10^{-14}$).  The
radial-transition detection routine numerically integrates an object
if its osculating orbit as determined on the outer step comes within
5\% of the transition zone.

\subsection{Single planet}

Here we consider a two-zone integrator with outer timestep 0.05 yr and
inner timestep 0.025 yr.  First we set the transition zone from 0.9
to 1.1 AU.  We place objects of masses ranging from 1 \ME to 10000 \ME
at semimajor axes between 0.8 and 1.2 AU, and consider the resulting
behaviour of the energy (in the one-planet case, a measure of the
variation in semimajor axis).  For objects which enter the transition
region, the surrogate Hamiltonian under which the planet is moving
contains terms from both the inner and outer zones.

\begin{figure}
\begin{center}
\includegraphics[width=85mm,height=73.6667mm]{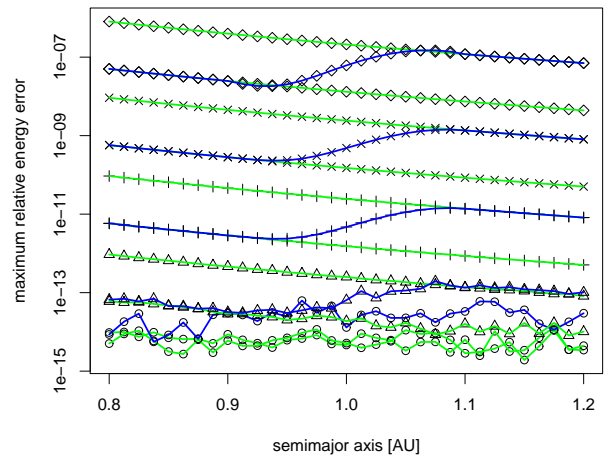}
\caption{Energy error as function of semimajor axis and object mass;
the DH runs for 0.025 and 0.05 yr timesteps are in green, the Naoko
runs are in blue.}
\label{singleplanet_relerr}
\end{center}
\end{figure}

Figure \ref{singleplanet_relerr} shows three points for each of the
five mass cases (1,10,100,1000, and 10000 \ME) at each initial
semimajor axis: one for the basic DH integrator
(eq.~\ref{eq:dh_basic_5_op}) with a timestep of 0.025 yr, one for
Naoko, and one for the DH integrator with a timestep of 0.05 yr.  The
errors generally decrease with increasing semimajor axis: for a fixed
timestep, at larger orbital radius there are more steps per orbit.  As
expected, for objects in the transition zone, the new code reports
errors which smoothly interpolate between those of the smaller and
larger DH runs, and become indistinguishable from the standard
algorithm outside the transition region.  In the 10000 \ME runs -- 3\%
of the mass of the Sun -- there are some slight deviations apparent at
the edges of the transition region.  The clean interpolation also
breaks down at very low mass for a different reason: the integration
error is dominated by the integration tolerance.

\begin{figure}
\begin{center}
\includegraphics[width=85mm,height=73.6667mm]{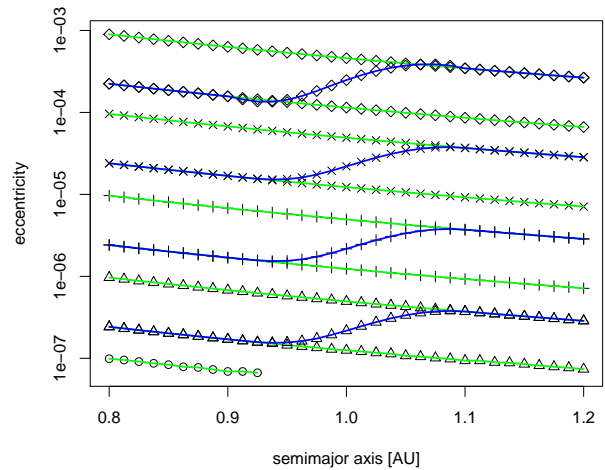}
\caption{Maximum eccentricity as function of semimajor axis and object
mass.  Most of the 1 \ME runs are not plotted, as their maximum $e$ was
0.}
\label{singleplanet_maxe}
\end{center}
\end{figure}

The situation is very similar for the eccentricities, as shown in
figure \ref{singleplanet_maxe}.  Again we see the new code
interpolating between the 0.05 and 0.025 yr DH runs.  This shows that
at least in a sufficiently smooth case, integrating an isolated object
on multiple timesteps need not introduce spurious energy or angular
momentum behaviour.

\begin{figure}
\begin{center}
\includegraphics[width=85mm,height=85mm]{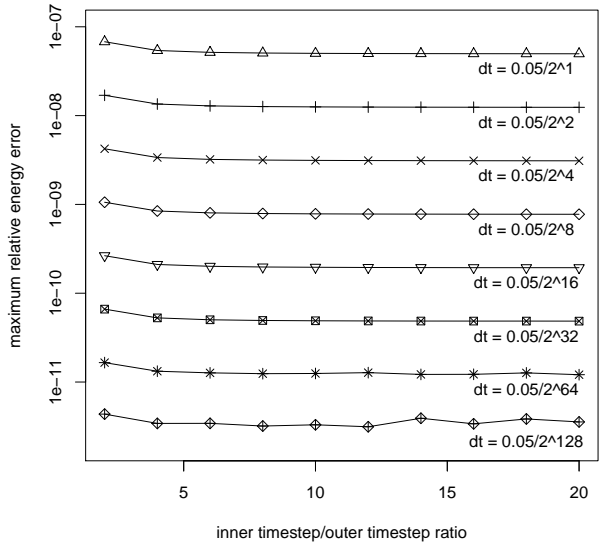}
\caption{Maximum relative energy error as function of timestep and
inner:outer timestep ratio for a 1 \ME object at 1 AU with $e=0.1$,
with transition zone from 0.95-1.05 AU.}
\label{conv1}
\end{center}
\end{figure}

Figure \ref{conv1} shows the relative energy error as a function both
of the timestep and the inner:outer timestep ratio, where the largest
outer timestep used was 0.05 yr, for a 1 \ME object at 1 AU with
$e=0.1$, with transition zone from 0.95-1.05 AU.  As expected, the
integrator behaves as a second-order algorithm.  Increasing the ratio
to take more inner steps per outer step improves the energy error very
little beyond 4:1, as the maximum error is controlled by the outer
step which is not changing.

Although this condition is necessary, it is clearly insufficient.
More realistic tests involve objects which repeatedly cross from the
inner zone to the outer zone through the transition zone during the
same orbit.

\begin{figure}
\begin{center}
\includegraphics[width=85mm,height=85mm]{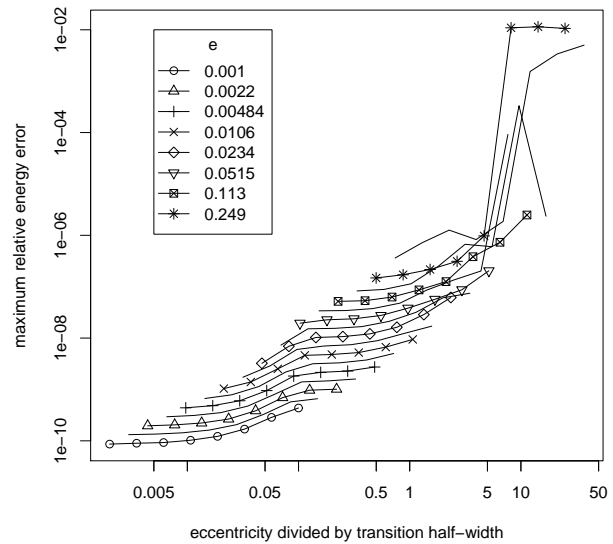}
\caption{Maximum relative energy error as a function of transition
width and eccentricity; the eccentricity increases between curves by a
factor of 1.48.}
\label{relmaxerr}
\end{center}
\end{figure}

Without loss of generality we placed a 1 \ME object in the centre of a
transition zone at 1 AU, and varied both the eccentricity $e$ and the
relative transition zone half-width $h$, where $h$ is defined so that
the zone extends from $a (1-h)$ to $a (1+h)$, in analogy with the
eccentricity.  The outer timestep was chosen to be 1/20th of the
orbital period, the eccentricities were varied from 0.001 to 0.37 (the
latter chosen because that corresponds to a perihelion for which 0.025
yr is 1/20th of the orbital period at that distance), and the
half-width was varied from 0.01 to 0.50 AU.  The resulting maximum
relative energy error is shown in figure \ref{relmaxerr} as a function
of the $e/h$ ratio; the contours correspond to constant $e$.

\begin{figure}
\begin{center}
\includegraphics[width=85mm,height=73.6667mm]{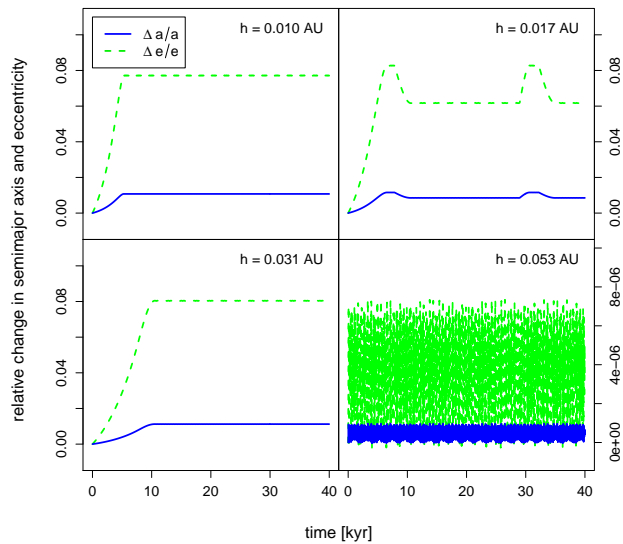}
\caption{Relative change in semimajor axis and eccentricity for
various transition half-widths $h$ at $e \simeq 0.25$; the three
half-widths below did not converge.}
\label{convexam}
\end{center}
\end{figure}

At small $e/h$ -- that is, when the transition zone is larger than or
comparable to the radial excursion per orbit -- the energy error is
well-behaved for a fixed $e$, and is only weakly dependent on $h$.  At
$e/h \gtrsim 10$, there are many integrations which do not converge.
Nonconvergence is easily recognized from the evolution of the
semimajor axis or the eccentricity: the planet begins to migrate away
from its original location until it finds a semistable configuration.
Examples are given in figure \ref{convexam} for $e \simeq 0.25$.  Not
only does the transition need to be sufficiently smooth along an
orbit, but for a run with a larger tolerance or a larger timestep, a
wider transition zone may be required so that the numerical integrator
can detect the transition (unless the code accepts hints about the
location of difficulties in the integrand).

\subsubsection{Varying the switch}
\label{section:vary_switch}

\begin{figure}
\begin{center}
\includegraphics[width=85mm,height=85mm]{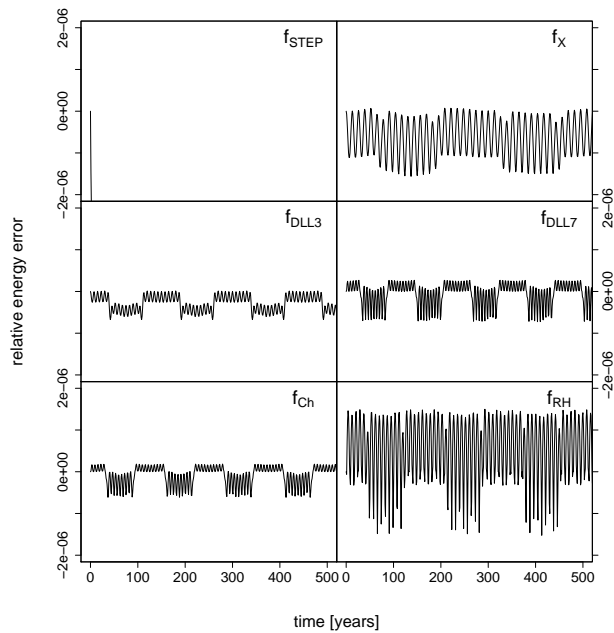}
\caption{Maximum relative energy error for single planet at 1 AU,
e=0.2, with transition zone 0.95-1.05 AU, for various switch functions.}
\label{varyswitch}
\end{center}
\end{figure}

We also explored the effects of alternate choices of transition
function.  We place a 1 \ME object at 1 AU with $e=0.2$, and a
transition zone extending from 0.95 to 1.05 AU with, with outer
timestep $0.05$ yr and inner timestep $0.025$ yr.  We then vary the
switch over the six functions described in \S\ref{section:trans}: the
step function; the switch f(x) = x; the DLL98 cubic; the DLL98 septic;
the Chambers switch; and the Rauch \& Holman switch.  Figure
\ref{varyswitch} shows the resulting relative energy errors: the
behaviours for the convergent integrations continue for millions of
orbits.

The step function fails immediately: the object migrates outward in
semimajor axis to $\sim\!1.00456$ AU over the first 2700 years and
then stops, corresponding to a maximum relative energy error of
$\sim\!0.0046$.  The remaining integrations all succeeded, with
$f_{\mathrm{DLL3}}$ showing the smallest energy error and
$f_{\mathrm{RH}}$ the largest.  We have found this is typical.  That
the crude $f_{\mathrm{X}}$ performs better than the smooth
$f_{\mathrm{RH}}$ can be understood by returning to figure
\ref{fig:trans}: $f_{\mathrm{RH}}$ is the smoothest function, but for
a fixed transition width it is also the one with the narrowest
effective transition (extending roughly from x=0.25 to x=0.75).

The execution times for the various switches (excluding the step
function) were generally comparable -- mostly agreeing within the
scatter -- except that the Rauch \& Holman switch was consistently the
slowest, and $f_{\mathrm{X}}$ the fastest.  The slower speed of the RH
switch is likely chiefly due to the evaluation of the trigonometric
function.  (Note that some versions of the common GCC C compiler will
occasionally refuse to inline functions for obscure reasons, leading
to strange profiling results.)

Thus we chose $f_{\mathrm{DLL3}}$ as the best compromise between
efficiency of evaluation and energy conservation.

\subsection{Multiple planets}

The above results are unsurprising, insofar as the use of smooth
transitions to move portions of the Hamiltonian from one part of the
step to another is familiar.  Of more concern is whether the reduction
in the number of force evaluations will lead to unacceptable
resolution of the angular momentum exchange between planets in
different zones.  We will concentrate on the interactions which are
likely to be the most sensitive to changes in sampling frequency,
namely resonant interactions.  Recall that we are willing to accept a
cruder approximation to the dynamics than we would ordinarily do, as
long as there are no clear integration failures.  Since one of our
interests is studying migration scenarios, in which resonances play a
significant role, it is important to verify that such behaviour is not
lost when using the multizone methods.

We will verify that resonant behaviour can persist when the objects
are in different zones, when one object is in a transition zone, and
when the objects migrate across a zone in resonance (coorbital or
otherwise).  In each simulation we have used the two-zone 0.05/0.025
yr 0.9-1.1 AU transition.  To make comparisons with the behaviour of
the DH integrator clearer we have disabled the encounter treatment for
the following sections (until \S\ref{sect:closeenc}).

\subsubsection{2:1 mean motion resonance}
\begin{figure}
\begin{center}
\includegraphics[width=85mm,height=73.6667mm]{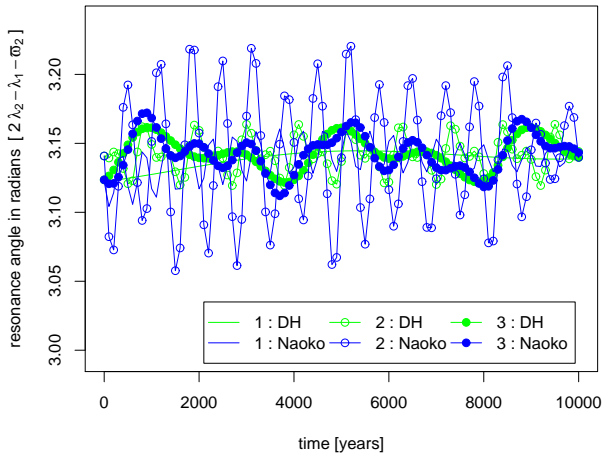} 
\caption{Evolution of resonance angles for pairs of objects in 2:1 mean
motion resonance under both DH (green) and Naoko (blue) integrators.
The three cases correspond to $a_1, a_2$ = (0.63 AU, 1.00 AU), (0.76
AU, 1.20 AU), and (1.00 AU, 1.59 AU).}
\label{2to1}
\end{center}
\end{figure}

We place two 2 \ME objects on cold $e\sim\!0.002, i \sim\! 0$ orbits
in the 2:1 mean motion resonance in three configurations: (1) $a_1 =
0.63$ AU, $a_2 = 1.00$ AU, where the outermost object is in the middle
of the transition region; (2) $a_1 = 0.76$ AU, $a_2 = 1.20$ AU, where
neither object is in the transition zone but the objects are in
different zones; and (3) $a_1 = 1.00$ AU, $a_2 = 1.59$ AU, where the
innermost object is in the transition zone.  (All semimajor axes are
approximate; the initial configurations were found by introducing a
slowly-decreasing dissipation into the DH algorithm.)  The evolution
of the resonance angles is plotted in figure \ref{2to1}.  In all cases
the resonance is preserved, and in the cases where one object was in a
transition zone the agreement is excellent between the Naoko results
and the results of a DH run with timestep 0.025 yr.  In the second
case -- in which neither object is in a transition zone -- the
libration is considerably larger in the Naoko run.  This is not
surprising: in the absence of dissipation the resonance is quite
sensitive to the initial conditions, and case 2 has the greatest
sudden change from DH to Naoko, as none of the force is being
evaluated on the higher (inner) frequency.  Even in this case, the
introduced libration is comparable to the differences in width between
different DH runs in which only the initial angles had been changed.
For our purposes it is sufficient that the objects remain resonant,
which they do.

\subsubsection{Co-orbitals}

\begin{figure}
\begin{center}
\includegraphics[width=85mm,height=73.6667mm]{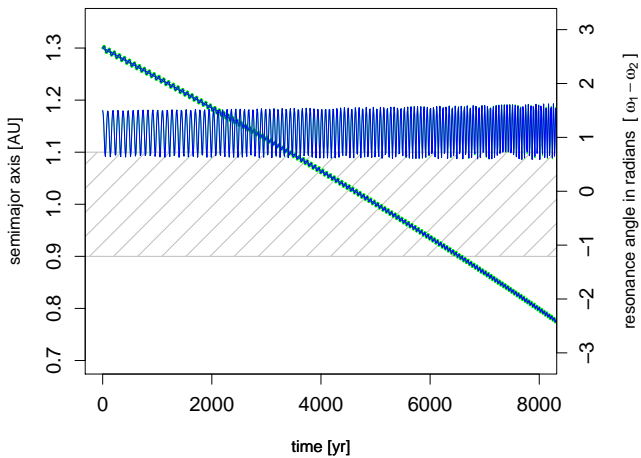} 
\caption{Semimajor axis and resonance angle for coorbital migration
pair crossing transition zone (the shaded area) for DH (green) and
Naoko (blue) integrators.}
\label{coorb}
\end{center}
\end{figure}

We place two objects -- one 10 \ME and one 1 \ME -- at 1.3 AU on
(osculating) circular and coplanar objects with mean anomalies
differing by 1.5 radians, and let them migrate inwards in the standard
Hayashi minimum mass solar nebula (MMSN) disc model \citep{hay1981}
with surface density $\simeq 1700$ g/cm$^2 (r/\mathrm{AU})^{-1.5}$.)
The migration was driven by a prescription for type I gas migration
\cite{tanakaward}, in which the motion is produced by the asymmetry
between the torques generated by the interior and exterior wakes of
the body \citep{goldtrem, ward}.  (The difference in mass between the
two test objects is to ensure that the lower-mass planet is being
carried along by the resonance and not merely migrating in tandem.)
Figure \ref{coorb} shows the evolution of the semimajor axis and the
resonant angle for both the DH and Naoko integrators.  Even when the
pair crosses the transition zone, the trajectories are barely
distinguishable.  It may seem counterintuitive that the algorithm
performs near-perfectly on coorbital resonances which depend
sensitively on the local potential and considerably worse on distant
resonances which might be expected to be more forgiving of sampling
errors.  However, the method's errors compared with the the standard
DH algorithm all involve {\em relative differences}, and in the case
of co-orbital objects on near-circular objects, these are minimal.

\subsubsection{Transition crossing}

\begin{figure}
\begin{center}
\includegraphics[width=85mm,height=73.6667mm]{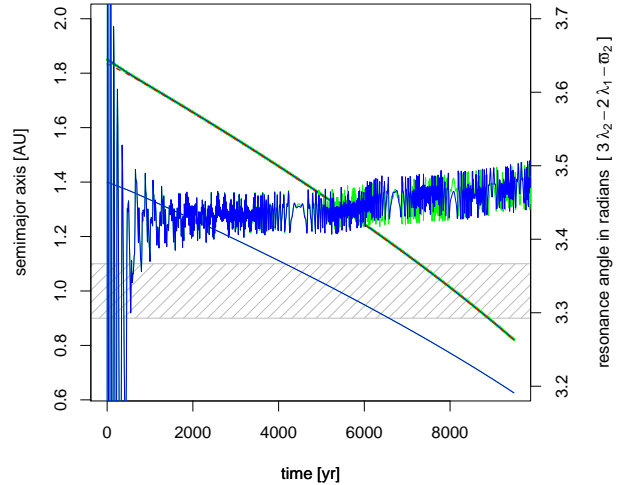} 
\caption{Semimajor axis and resonance angle for pair of objects in 3:2
resonance crossing transition zone (the shaded area) for DH (green)
and Naoko (blue) integrators.}
\label{3to2}
\end{center}
\end{figure}

A more rigorous test comes from seeing whether resonance can be
maintained when the objects are under different -- and varying --
terms in the Hamiltonian.  We place a 10 \ME planet at 1.40 AU and a
20 \ME planet at 1.85 AU, just outside the 3:2 resonance at
$\sim\!1.83$AU, and turn on type I migration in the MMSN for them
both.  The results are shown in fig. \ref{3to2}.  The 20 \ME planet
migrates faster than the 10 \ME planet and catches up to the resonance
within the first $\sim\!1000$ yrs.  The evolution under DH of 0.025 yr
and Naoko with 0.025/0.050 yr is almost identical until the inner
object enters the transition zone; that Naoko is using a timestep
twice as large makes little difference.  Even after the objects enter
the transition region, the planets remain in the 3:2, and the
libration width is comparable between the two runs.

\subsubsection{Transition crossing: varying timestep ratios}

\begin{figure}
\begin{center}
\includegraphics[width=85mm,height=73.6667mm]{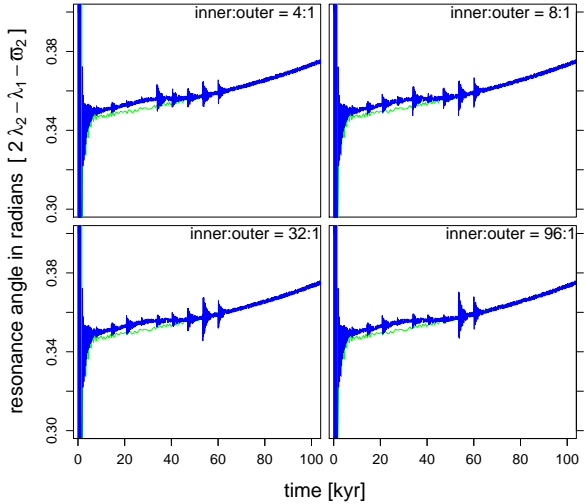} 
\caption{Evolution of resonance angle for 2:1 pair crossing a
transition region as the inner:outer ratio is varied; the thin line is
the reference single-timestep behaviour}
\label{vary_M}
\end{center}
\end{figure}

It is of interest to see how the maintenance of the resonance varies
as the inner:outer timestep ratio is varied.  We place terms in the
Hamiltonian.  We place a 1 \ME planet at 0.77 AU and a 10 \ME planet
at 1.23 AU, just outside the 2:1 resonance, and again turn on MMSN
type I migration.  We fix the outer timestep at 0.05 yr, and change
the number of inner steps per outer step, while keeping the transition
zone from 0.90 AU to 1.10 AU in each run.  Figure \ref{vary_M} shows
the resulting evolution of the resonance angle.  The use of multiple
timescales for the forces introduces spurious libration spikes (with
corresponding behaviour in the eccentricities), with the libration
width during the spikes generally increasing with the timestep ratio.
However, in all cases, from 2:1 through 96:1, the integration error
failed to break the resonance during the transition, and the pair
returned to the correct trajectory after the outer object passed
through the transition region.

\subsubsection{Transition crossing: effect of force frequency}

\begin{figure}
\begin{center}
\includegraphics[width=85mm,height=73.6667mm]{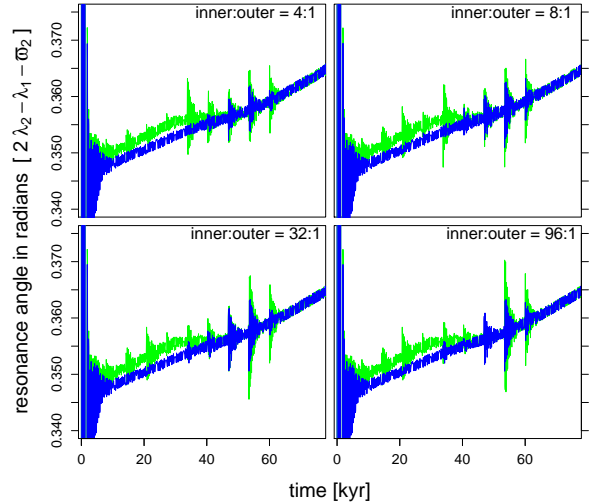}
\caption{Evolution of resonance angle for 2:1 pair crossing for
different transition region as the inner:outer ratio is varied; the
heavier line corresponds to the high frequency force evaluation, and
the lighter line to the standard low frequency evaluation of figure
\ref{vary_M}.}
\label{vary_M_highfreq}
\end{center}
\end{figure}

As mentioned in \S\ref{section:naoko1}, although we have chosen to
evaluate the forces on the outermost timestep, one can vary this
frequency.  We repeat the test of the previous section but with the
forces always evaluated on the inner timestep (and accordingly with no
need for any transition function applied to the kick operators), in
which the only active transitions involve the timestep on which the
object is being drifted.  The resulting evolution of the resonance
angle is shown in figure \ref{vary_M_highfreq}.  We see that using the
fixed high-frequency forces results in a considerable qualitative
improvement, although the multiple timescales used for the Kepler
drift still generate libration spikes.  This serves to bound the
likely improvement we can imagine by increasing the force frequency,
which is very costly.  Recall that the formation scenarios we seek to
study using the new method involve large particle numbers where the
execution time is dominated by the force calculation -- a quite
different regime from the solar system modelling of \cite{saha94}.

\subsubsection{Convergence}
\label{2ressect}

\begin{figure}
\begin{center}
\includegraphics[width=85mm,height=85mm]{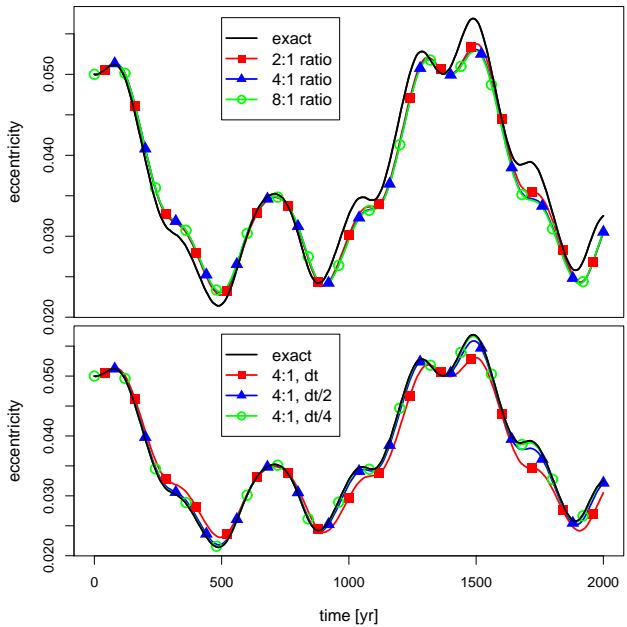} 
\caption{Eccentricity evolution for inner object in near-resonant pair
described in \S\ref{2ressect} for various timesteps and inner:outer
timestep ratios; the system does not converge to the true path as the
ratio is increased but does as the timestep is decreased.}
\label{conv2}
\end{center}
\end{figure}

It is important to recognize that by changing the time sampling as we
have done, we are unavoidably changing the dynamics of the system.
Since the outer timestep remains fixed as we increase the inner
timestep, and with it some fraction of the force evaluation, we are
modifying the effective interaction between the inner and outer
objects.  For example, we consider a pair of 1 \ME objects near a
nominal 2:1 resonance at $\sim\!0.79$ AU and $1.25$ AU and vary the
timestep and inner:outer ratio.  The original outer timestep was 0.05
yr.  Figure \ref{conv2} shows the resulting change in eccentricity
evolution for the inner object as both the inner:outer timestep ratio
and the outer timestep are varied (the outer object shows similar
behaviour.)  As the ratio is increased in the top section of the
figure -- i.e. as the inner timestep is decreased -- the eccentricity
evolution does converge, but it does not converge to the correct
trajectory, although the relative differences in this case are not
large.  For a fixed timestep ratio, of course, as the outer timestep
is decreased, the evolution converges as it should.

\subsubsection{Close encounters}
\label{sect:closeenc}

As mentioned previously, when two objects are in the same zone the
close encounters are treated as in the Chambers algorithm, and 
we disallow encounters between objects which are not in
neighbouring zones.  Since it is known that accurate resolution of
close encounters is quite sensitive to the effective Hamiltonian under
which it is being integrated (DLL98, \S4), it is reasonable to expect
poor behaviour for objects undergoing close encounters in a transition
region where it is undergoing drifts of different duration.  A
decrease in accuracy may be tolerable at these few locations in a
simulation, but the objects cannot suffer dramatic instabilities.

\begin{figure}
\begin{center}
\includegraphics[width=85mm,height=73.6667mm]{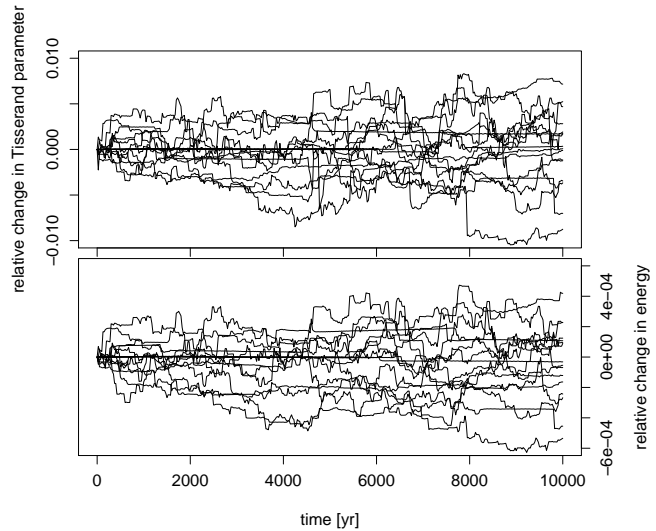}
\caption{Energy conservation and variation of Tisserand parameter
during close encounters described in \S\ref{sect:closeenc}.}
\label{closeenc}
\end{center}
\end{figure}

We use the 0.9-1.1 AU, dt=0.05 yr/0.025 yr transition.  To generate
frequently-encountering systems, we placed a 5 \ME planet at 1.00 AU,
in the middle of the transition region, on a circular orbit and a
protoplanet of 0.1 \ME at 1.5 AU with $e=0.333$ on a coplanar orbit,
set all angles to 0 and varied only the mean anomaly from 0 to $2
\pi$.  After removing runs in which the smaller body did not undergo
an encounter, would have suffered a potential merger, or (in one case)
became sufficiently eccentric that it escaped the integration region,
16 runs remained.  The relative change in the energy and the Tisserand
parameter $T = 1/2a +\sqrt{a (1-e^2)}$ (\citealt*{murray} \S 3.4) is
plotted in figure \ref{closeenc}.  The RMS of the maximum changes in E
is $\sim\!0.0004$, and that of T $\sim\!0.01$; the horizontal segments
correspond to periods without encounters, where the conservation
properties revert to the standard non-encountering behaviour.  For
comparison, the equivalent RMS maximum changes for our reference SyMBA
implementation with timestep 0.025 yr -- with mergers disallowed, to
make the integration even more difficult -- were $\sim\!3\,\cdot\, 10^{-6}$ and
$\sim\!0.003$ for E and T, respectively.  It is clear that the
repeated close encounters in the transition zone cause a considerable
decrease in accuracy, especially in the energy, but the error growth
should be tolerable.

The Bulirsch-Stoer numerical integrator we use often fails, and
reports its failure, at high precision for very close encounters.
This is typically not a problem in practice, as the separations at
which the integrator fails are usually much smaller than the distances
at which we would merge the two bodies, but is an issue for
small-radius dynamical problems.  The same failure mode exists when no
transition zones are involved and the symplectic algorithm is one
resembling that of Chambers.

\subsection{Full example: short-period Neptunes}

As a case study, we consider the problem of forming short-period
Neptunes, which are particularly mysterious.  It is clear they did not
form at the small orbital radii at which they are currently found, and
so it is likely they have migrated in from further out.  However,
their masses -- $10$\ME and higher -- are unexpected.  They are large
enough that the timescale for type I migration to drag them into the
Sun is considerably shorter than the formation timescale, but too
small for some plausible mechanisms for suppressing type I migration
to be effective (such as the opening of a gap in the disc at the
transition to type II migration.)

There are two main scenarios present in the literature for forming
these objects.  The first, due to \cite{alib}, concentrates on
sophisticated gas physics and follows the evolution of effectively
isolated cores through the disc as they grow via planetesimal
accretion and migrate, and finally have their atmospheric mass reduced
after arriving in the short-period region by evaporation.  The second,
due to \cite{terq}, takes a more traditional N-body approach, and
follows the evolution of a small number of large embryos in the inner
region.  Neither of these are easy to reconcile with the standard
oligarchic growth model for core formation \citep{ko98}: the first
requires core formation to be suppressed everywhere in the disc except
at a few special locations, and the second uses initial conditions for
the mass distribution which it is difficult to recover
self-consistently from earlier stages of oligarchic evolution.

It is of considerable interest whether the standard model can generate
anything resembling the observed planets -- if it cannot, then we have
strong evidence that we are missing important physics -- and the new
method can address this question.  We will merely sketch the
application here; forthcoming work will describe our results in more
detail.  We apply the two-stage approach of \cite{mcn05}, which was
based on that of \cite{edolig}, using semianalytic approximations of
oligarchic evolution to treat the early evolution of the disc and then
generating an N-body realization when the number of N-body particles
needed drops to practical levels.  We consider various disc models
(all resembling the minimum mass model of \citealt*{hay1981}) and
different migration efficiencies.

\begin{figure}
\begin{center}
\includegraphics[width=85mm,height=109.481mm]{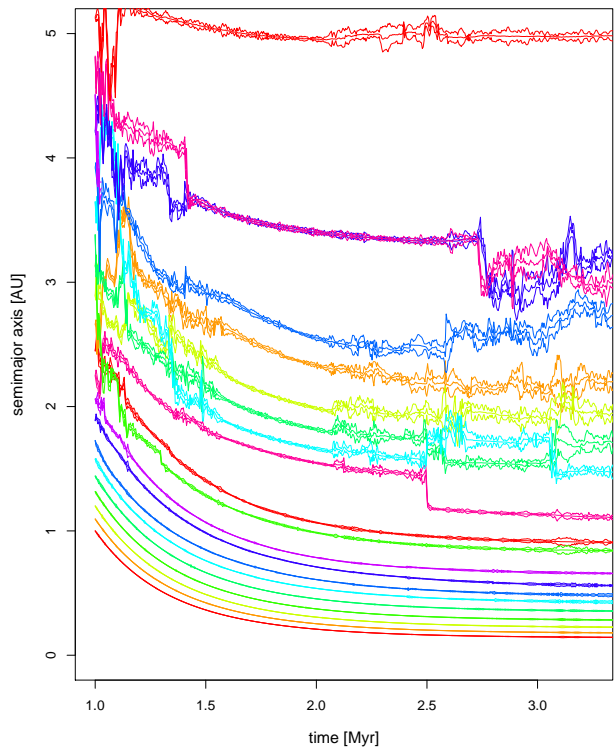}
\caption{Embryo evolution with time for a simulation of a disc 3 times
the minimum mass with $\Sigma \propto r^{-0.5}$, with disc dissipation
e-folding time of 0.5 Myr.  For each embryo lines corresponding to
semimajor axis, perihelion distance, and aphelion distance are drawn.
\label{example1_029}}
\end{center}
\end{figure}

An example of a low-resolution run is shown in figure
\ref{example1_029} which employed 21 embryos and 212 planetesimals.
The semianalytic model was run to 1 Myr for a disc with surface
density proportional to $r^{-0.5}$ of $\sim\!3$ times the standard
Hayashi mass without a snow line, with nominal type I migration and
aerodynamic drag (for a 1 km planetesimal size), where the e-folding
time for the decay of the gas disc was 0.5 Myr.  A section of the
system ranging from 1.0 AU to 5.1 AU was then evolved under the new
N-body code until the gas was mostly absent.  As a consequence of the
semianalytic model predicting that the innermost part of the disc had
gone to oligarchic completion, the embryos inside of 2.0 AU migrate
inwards smoothly in tandem: there is nothing to perturb them.  This
results in $10.3$ \ME of embryo material inside of 1 AU when the gas
vanishes in an apparently stable configuration.  Further out, the
system is still undergoing chaotic evolution.  In a more realistic
situation, in which there were still planetesimals around in the inner
regions to perturb the embryos, they might merge, producing larger
planets; or, if they merge too early, they could migrate into the Sun.

\begin{figure}
\begin{center}
\includegraphics[width=85mm,height=109.481mm]{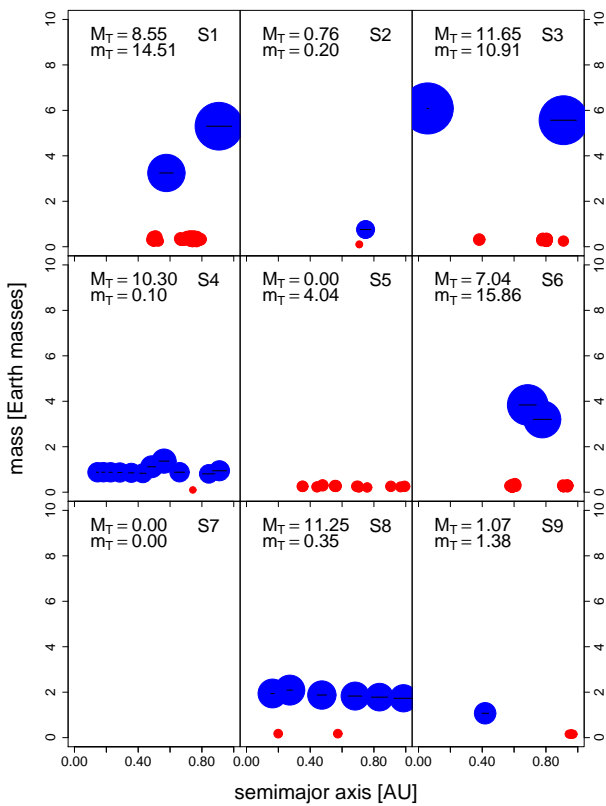}
\caption{Mass versus semimajor axis snapshots at T=3.25 Myr for
various runs; the larger blue circles represent embryos, and the
smaller red circles represent planetesimals.  The total mass in
embryos ($M_T$) and planetesimals ($m_T$) in the region inside 1 AU is
given in the upper left.
\label{grid9}}
\end{center}
\end{figure}

Figure \ref{grid9} presents examples of the kinds of interior
configurations our toy simulations produce; all are plotted at t=3.25
Myr.  Typical results include failures in which no or almost no mass
is left in the interior region (S2, S5, S7, S9); cases where the total
mass is interesting from the perspective of hot Neptune formation but
the embryo mass is spread out over a large number of embryos instead
of being concentrated in one or two objects (S4, S8); and cases where
the total mass is appealing and there are only two embryos of multiple
Earth masses (S1, S3, S6) but where too much of the mass is locked up
in planetesimals, possibly as a consequence of the low resolution.
(Planetesimals do not self-interact in these simulations.)  Even these
crude simulations are useful to estimate whether the conditions are
appropriate for larger integrations.  We will present results from
higher resolution simulations in a forthcoming publication.

\section{Discussion}
\label{section:disc}

The observed behaviour of the method (including on tests not presented
above) is compatible with the principles that (1) parts of the
Hamiltonian can be moved between operators with different timesteps if
the transition is sufficiently smooth, (2) reversibility is more
important than accuracy for secular energy conservation, and (3)
decreased time resolution is not catastrophic for dynamics on much
longer timescales.  We emphasize that we do not claim, and have not
shown, that the integrators are in any sense optimal.  (Some evidence
suggests that a drift-kick-drift splitting can be preferable to the
kick-drift-kick splitting used here, for example; see
\citealt{wisholtou}.)  To use a loose analogy from the history of
close encounter treatment, this approach probably lies somewhere
between the hybrid methods of \cite{ld94} and the mature methods of
DLL98.  As always, one must bear the approximations being used in mind
-- small effects can accumulate to cause dramatic changes on secular
timescales in ways not always easy to predict.  The approximations
used here have the virtue of being different from those used
previously.  Given the decrease in force accuracy involved, it may be
best to think of the use of these methods as one would a reversible,
Kepler-adapted treecode: useful for studying certain formation
problems, and generally informative in a statistical sense, but
inappropriate for cases where detailed dynamics are important.
Potential users are advised to think carefully about whether the
approximations are suited for their problems, as there are far more
ways for an integration to fail in the multi-zone case than there are
in standard SyMBA-style integrators.

There are several potential applications and directions for
generalization we have not considered here.  As discussed in section
\ref{section:close_enc_with_Sun}, \cite{ld00} modified the Chambers
approach to handle objects which occasionally require very small
timesteps, at the price of integrating the entire system numerically
when such an approach occurs.  The new method succeeds in decoupling
such objects, and so may be practical in some cases for which their
method is inapplicable (such as when ejections of low-mass objects are
common).  It may also be useful in studying planet formation in close
binaries, for which some symplectic methods for dealing with these
systems already exist \citep{chamquint}.

\cite{saha94} develop `symplectic interpolation' methods to improve
the accuracy of their integrator, by using a symplectic prediction of
the effects of $\HKep$ to better synchronize the force calculations.
This would probably work here as well, in a formal sense, but it is
unlikely to provide dramatic improvements, and the errors in energy
and angular momentum conservation introduced by the method should be
tolerable for most formation problems as they stand.

\section{Conclusions}
\label{section:conc}

We have presented a new integration method which preserves most of the
speed and conservation properties of standard close-encountering
symplectic integrators for planetary dynamics, but introduces radial
zones between which the drift and force evaluation timesteps can vary.
This allows for a new tradeoff between the accuracy of the integration
and the speed, one which is inappropriate for precise dynamical
studies but may be of use for investigating other scenarios.  We
expect it to be useful for approximate N-body modelling of planetary
formation problems where (1) there is a wide range of orbital
velocities, (2) the majority of the material needing particle
representation is in the regions with longer orbital periods, and (3)
the system remains sufficiently cold that the transition regions
between radial zones can be kept reasonably small.  Migration
scenarios of the formation of hot exoplanets satisfy all these
constraints, and we are currently exploring such applications.

\section*{Acknowledgments}

DSM thanks Martin Duncan and Paul Wiegert for useful discussions, and
thank the anonymous referee for a careful review.  The authors
gratefully acknowledge the support of SFTC grant PP/D002265/1.  The
simulations presented in this paper were performed using the QMUL HPC
facilities.

\label{lastpage}

\end{document}